\documentclass[superscriptaddress, twocolumn,showkeys,
amssymb,amsmath,nobibnotes,aps,prd,
nofootinbib]{revtex4-2}
\pdfoutput=1
\usepackage{svg}
\usepackage{graphicx,subfigure,bm,color,psfrag,hyperref}
\usepackage[export]{adjustbox}
\usepackage{amsfonts}
\usepackage{lipsum}
\usepackage{mathtools}
\usepackage{verbatim}
\usepackage[normalem]{ulem}
\usepackage[dvipsnames]{xcolor}
\hypersetup{colorlinks,linkcolor={blue},citecolor={red},urlcolor={greenW}}  
\definecolor{greenW}{rgb}{0.0, 0.55, 0.1} 
\usepackage[inline]{enumitem}
\usepackage{booktabs}
\usepackage{footmisc}

\usepackage{multirow}
\usepackage{tabularx}
\usepackage{array}
\newcolumntype{C}[1]{>{\centering\arraybackslash}p{#1}}

\begin{document}

\title{Reconstructing dark energy with fewer assumptions}

\author{Daniel A. Kessler}
\email{d.kessler@sheffield.ac.uk}
\affiliation{School of Mathematical and Physical Sciences, University of Sheffield, Hounsfield Road, Sheffield S3 7RH, United Kingdom}

\author{Eleonora Di Valentino}
\affiliation{School of Mathematical and Physical Sciences, University of Sheffield, Hounsfield Road, Sheffield S3 7RH, United Kingdom}

\author{Luis A. Escamilla}
\affiliation{Department of Physics, Istanbul Technical University, 34469 Maslak, Istanbul, Turkey}

\author{Dragan Huterer}
\affiliation{Department of Physics and Leinweber Institute for Theoretical Physics,
University of Michigan, 450 Church St, Ann Arbor, MI 48109}


\begin{abstract}

We perform minimalistic reconstructions of the dark energy density and equation of state using late-time distance measurements. Our methodology avoids assumptions that correlate the values of these functions over time and instead yields their approximate average evolution within seven redshift bins from $z=0$ to $z=4.2$. Constraints are obtained using combinations of baryon acoustic oscillation measurements from the Dark Energy Spectroscopic Instrument (DESI) and the completed Sloan Digital Sky Survey (SDSS), alongside Type~Ia supernovae measurements from Pantheon+ and the latest recalibrated samples, Union3.1 and DES-Dovekie. Only an acoustic scale prior is included from the Cosmic Microwave Background so that our results are insensitive to the possible matter density tension between early and late-time probes. All combinations yield consistent reconstructed histories: a dark energy density that rises to a local maximum before decreasing at late times and an equation of state with two apparent oscillations around the cosmological constant limit. Both functions tentatively suggest a phantom crossing in the equation of state around $z\sim0.6$--$0.8$. These patterns are robust to numerous parameter extensions, such as freely varying spatial curvature and neutrino mass, and they persist in the uncorrelated amplitudes obtained through localized principal component analysis. Deviations from $\Lambda$CDM in individual bins reach a maximum significance of $\sim2.6$--$3\sigma$, while the total chi-square difference between the reconstructions and this model provides up to $\sim2\sigma$ support for the seven additional parameters in the reconstructions. As these significances remain moderate, our main result is the level of consistency between combinations of the most widely used background-level observations. Our results suggest that the dark energy evolution signal is a persistent feature of the data and that it cannot be explained solely by fluctuations or systematics in individual measurements.

\end{abstract}
\maketitle

\section{Introduction}

Local distance measurements from Type Ia supernovae (SNe~Ia)~\cite{Riess:1998cb,Perlmutter:1998np,Brout:2022vxf} and baryon acoustic oscillations (BAO)~\cite{eBOSS:2020yzd,DESI:2025zgx}, alongside cosmic microwave background (CMB) anisotropies~\cite{Planck:2018vyg,ACT:2025fju,SPT-3G:2025bzu} and large-scale structure probes~\cite{Wright:2025xka,DES:2026fyc}, provide compelling evidence that the Universe is currently undergoing accelerated expansion. Under the assumptions of homogeneity and isotropy on large scales, this acceleration requires either a modification of gravity or a matter species whose pressure is less than $-1/3$ of its energy density, termed \textit{dark energy}~\cite{Huterer:2017buf}. The simplest realization is the cosmological constant, $\Lambda$, which is naturally present in general relativity and, together with cold dark matter (CDM), forms the basis of the standard cosmological model, $\Lambda$CDM.

Despite the remarkable phenomenological success of $\Lambda$CDM, the physical origin of the cosmological constant remains unclear. Its observed energy scale is difficult to reconcile with expectations from quantum field theory, and its apparent coincidence with the present matter density has long motivated searches for more general forms of dark energy~\cite{Weinberg:1988cp,Carroll:2000fy,Peebles:2002gy,Copeland:2006wr,Martin:2012bt}. A simple phenomenological test consists of promoting the dark energy equation of state, $w_{\rm DE}=p_{\rm DE}/\rho_{\rm DE}$, to a free constant parameter. In this case, current data generally yield constraints close to the cosmological constant limit, $w_{\rm DE}=-1$~\cite{Planck:2018vyg,Aubourg:2014yra,eBOSS:2020yzd,Brout:2022vxf,DES:2024tys,DESI:2025zgx,Escamilla:2023oce}. However, a richer phenomenology is possible if the equation of state is allowed to evolve with time. The most widely used two-parameter description is the Chevallier-Polarski-Linder (CPL) parametrization, $w_{\rm DE}(a)=w_0+w_a(1-a)$, where $a$ is the scale factor~\cite{Chevallier:2000qy,Linder:2002et}. While originally proposed to mimic the behavior of scalar field models~\cite{Linder:2002et}, this parametrization is often interpreted as a first-order expansion in the scale factor around the present epoch, $1-a=0$. In this framework, $\Lambda$CDM is recovered when $(w_0,w_a)=(-1,0)$.

Recent data have shown hints of deviation from $(w_0,w_a)=(-1,0)$, significantly revitalizing interest in dark energy evolution. Statistically notable $\sim 2\sigma$ deviations were found when SNe~Ia measurements from the Pantheon+, Union3, and the Dark Energy Survey (DES) collaborations were combined with CMB anisotropies from \textit{Planck} and BAO observations from the Sloan Digital Sky Survey (SDSS)~\cite{Brout:2022vxf,Rubin:2023ovl,DES:2024tys}. The latest BAO measurements from the Dark Energy Spectroscopic Instrument (DESI), based on the first three years of observations, strengthened this preference in several combinations with CMB and SNe~Ia data, typically preferring the $w_0>-1$ and $w_a<0$ quadrant~\cite{DESI:2025zgx,Giare:2025pzu}. While these deviations from $\Lambda$CDM previously depended appreciably on the SNe~Ia compilation used, the recalibrated DES-Dovekie~\cite{Popovic:2025glk,DES:2025sig}, Union3.1~\cite{Hoyt:2026fve,Rubin:2026qdt}, and Pantheon+ compilations now contribute to consistent preferences for dark energy beyond $\Lambda$ at around $3.2$ to $3.4\sigma$ significance~\cite{DES:2025sig,Hoyt:2026fve}.

These results have generated considerable discussion~\cite{Bhattacharya:2024hep,Carloni:2024zpl,Chan-GyungPark:2024mlx,Cortes:2024lgw,Gialamas:2024lyw,Giare:2024gpk,Giare:2024oil,Jiang:2024xnu,Luongo:2024fww,Najafi:2024qzm,Reboucas:2024smm,RoyChoudhury:2024wri,Shlivko:2024llw,Tada:2024znt,Wang:2024dka,Ye:2024ywg,Cheng:2025hug,Cheng:2025lod,Cheng:2025yue,Chudaykin:2025aux,Chudaykin:2025lww,Cortes:2025joz,Fazzari:2025lzd,Herold:2025hkb,Ishak:2025cay,Keeley:2025rlg,Kessler:2025kju,Lee:2025pzo,Li:2025vuh,Ozulker:2025ehg,RoyChoudhury:2025dhe,Santos:2025wiv,Scherer:2025esj,Smith:2025icl,Specogna:2025guo,Wolf:2025jlc,Gokcen:2026pkq,Najafi:2026kxs,Yang:2026yaq}, but their physical interpretation remains unsettled. One active question is how the assumed dark energy model affects the evidence and phenomenology inferred from the data. While low-dimensional parametrizations like CPL are attractive for their ability to compress the inference problem into a limited number of interpretable parameters, they may obscure whether the data prefer genuinely smooth trends or, \textit{e.g.}, a richer structure of localized deviations. For this reason, it is useful to complement these parametrizations with more flexible dark energy reconstructions capable of resolving the data's preferences on smaller redshift scales~\cite{DESI:2024aqx,DESI:2025fii,Berti:2025phi,Li:2025ula,Zhang:2025bmk,DESI:2025wyn}.

In this work, we separately parametrize the normalized dark energy density, $f_{\rm DE}(z)\equiv {\rho_{\rm DE}(z)}/{\rho_{\rm DE}(0)}$, and the dark energy equation of state, $w_{\rm DE}(z)$, by piecewise-constant functions of redshift, approximately inferring the average evolution of these functions within seven redshift bins spanning $z=0$ to $z=4.2$. These two reconstructions are complementary. The energy density is more directly connected to the expansion history and, therefore, more tightly constrained by distance measurements. By contrast, the equation of state provides clearer indications of dark energy microphysics but is separated from distance measurements by an additional redshift integral, complicating its reconstruction. Constraining both quantities provides tighter constraints on deviations from $\Lambda$CDM through $f_{\rm DE}(z)$ and more intuitive, but less precise, guidance on the microphysical models consistent with the data through $w_{\rm DE}(z)$. 

These reconstructions are constrained by late-time distance measurements: BAO measurements from DESI~\cite{DESI:2025zgx} and the completed SDSS survey~\cite{eBOSS:2020yzd}, alongside SNe~Ia observations from Pantheon+~\cite{Brout:2022vxf} and the latest DES-Dovekie~\cite{Popovic:2025glk,DES:2025sig} and Union3.1 samples~\cite{Hoyt:2026fve}.\footnote{At the time of writing, the publicly available Pantheon+ data have not been corrected for the potential systematics identified in Refs.~\cite{DES:2025sig,Hoyt:2026fve}, so results including this dataset should be treated with some caution.} To calibrate the BAO measurements, we include a Big Bang Nucleosynthesis (BBN) prior on the physical baryon density. The only CMB information used in our baseline analysis is a conservative prior on the acoustic scale, $\theta_*$, which is robust to many cosmological assumptions~\cite{Vonlanthen:2010cd,Planck:2018vyg,Lemos:2023xhs}. Omitting a CMB prior on the physical cold dark matter density, a more model-dependent quantity, allows us to focus on the information contained in late-time distances, independent of the potentially confounding effects of the ``$\Omega_m$~tension''~\cite{Tang:2024lmo,Shlivko:2026jxa}. Our analysis is furthermore independent of concerns surrounding the optical depth, $\tau$, as measured by low-$\ell$ \textit{Planck} polarization data~\cite{Giare:2023ejv,Forconi:2023izg}, and its interplay with the evolving dark energy signal~\cite{Sailer:2025lxj,Jhaveri:2025neg}.

Our reconstruction methodology was carefully chosen to minimize assumptions and to remain as data-driven as possible. The only arbitrary choice entering piecewise-constant reconstructions is the division between redshift bins. We align our bins with the redshift ranges of the DESI-BAO tracers between $0.1\leq z<4.2$, including one additional bin for $z<0.1$ to account for the many SNe~Ia measurements at these low redshifts. This choice naturally ensures that BAO measurements are evenly distributed across the bins and that the reconstructions are fully robust to uncertainties in the DESI-BAO measurements' effective redshifts. Including SNe~Ia observations mitigates the risk of overfitting the statistical noise in the BAO measurements alone.

This paper is organized as follows. In Sec.~\ref{sec:methodology}, we describe our cosmological background assumptions, dark energy reconstruction and parameter inference methods, and the statistics used to quantify deviations from $\Lambda$CDM and internal consistency. In Sec.~\ref{sec:datasets}, we summarize the datasets considered. Sec.~\ref{sec:results} contains our main results: reconstructions of $f_{\rm DE}(z)$ and $w_{\rm DE}(z)$ from different BAO and SNe~Ia combinations. In Sec.~\ref{sec:extensions-robustness}, we assess the impact of including additional CMB information, of varying parameters held fixed in our baseline analysis (the spatial curvature, neutrino mass sum, and effective number of relativistic species), and of relaxing our assumptions regarding the sound horizon at baryon drag. In Sec.~\ref{sec:model-comparison}, we compare the reconstructions with $\Lambda$CDM expectations using both local and global statistics, and assess their internal consistency using the suspiciousness statistic. We apply localized principal component analysis in Sec.~\ref{sec:lpca} to identify the number of uncorrelated deviations from $\Lambda$CDM, and we present our conclusions in Sec.~\ref{sec:conclusions}.

\section{\label{sec:methodology}Methodology}

\subsection{\label{sec:background}Background}

In the $\Lambda$CDM model and its minimal extensions, the matter content of the Universe determines the expansion history through the relationship
\begin{multline}
    H(z)=H_0\bigg[\Omega_{\rm bc}(1+z)^{3}+\Omega_\gamma(1+z)^{4}\\+\Omega_k(1+z)^{2}+\Omega_\nu \frac{\rho_\nu(z)}{\rho_{\nu,0}}+\Omega_{\rm DE}\frac{\rho_{\rm DE}(z)}{\rho_{\rm DE,0}}\bigg]^{1/2}
    \label{eq:Hz}
\end{multline}
between the Hubble parameter, $H(z)=(\mathrm d a/\mathrm d t)/a$, and present-day energy densities relative to the critical density, $\Omega_i =8\pi G\rho_{i,0}/3H_0^2\equiv \rho_{i,0}/\rho_{\rm crit,0}$, where $H_0$ is the Hubble constant, and the index $i$ refers to baryons and cold dark matter ($\rm{bc}$), relativistic matter ($\gamma$), curvature ($k$), neutrinos ($\nu$), and dark energy (DE). In Eq.~\eqref{eq:Hz}, the latter two densities are written as general redshift-dependent functions, allowing for the transition between the relativistic and non-relativistic behavior of massive neutrinos and for a general dark energy evolution.

The expansion history then determines the transverse comoving distance,
\begin{equation}
    D_{\rm M}(z)=\frac{c}{H_0\sqrt{\Omega_k}}\sinh\left[\sqrt{\Omega_k}\int_0^z\frac{dz'}{H(z')/H_0}\right]\,,
\end{equation}
and the Hubble distance, $D_{\rm H}(z)=c/H(z)$. Baryon acoustic oscillations constrain the angular distances $D_{\rm M}/r_{\rm d}$ and $D_{\rm H}/r_{\rm d}$ with respect to the sound horizon at baryon drag, $r_{\rm d}$, as well as their geometric average, $D_{\rm V}/r_{\rm d}=(z\,D_{\rm M}^2\,D_{\rm H})^{1/3}/r_{\rm d}$. Type~Ia supernovae constrain the luminosity distance, $D_{\rm L}(z)$, through the distance modulus, $\mu$, after calibrating or marginalizing over the absolute magnitude. These distances are traditionally related by the Etherington distance-duality equation\footnote{See Refs.~\cite{Tang:2024zkc,Teixeira:2025czm,Afroz:2025iwo,Kanodia:2025jqh,Hu:2026yda} for recent examinations of this assumption.} $D_{\rm L}(z)=(1+z)D_{\rm M}(z)$ and, in a spatially flat Universe, by the integral relation $D_{\rm M}(z)=\int_0^z\mathrm{d}z'\,D_{\rm H}(z')$. Finally, the CMB constrains the angular sound horizon at recombination, $\theta_*=r_*/D_{\rm M}(z_*)$, where $r_*\simeq r_{\rm d}/1.02$ is the physical sound horizon at the redshift of recombination, $z_*$.

To efficiently calculate these background observables with transparent assumptions and approximations, we wrote a custom theory module for \texttt{Cobaya}~\cite{Torrado:2020dgo}, whose relevant analysis choices are described below.

\subsubsection{Neutrinos}

To obtain the neutrino contribution to the expansion history, we use the approximate expression from Ref.~\cite{WMAP:2010qai},
\begin{multline}
    \Omega_\nu \frac{\rho_\nu(z)}{\rho_{\nu,0}}=
    \left(\frac{7}{8}\right)
    \left(\frac{4}{11}\right)^{4/3}
    \left(\frac{N_{\rm eff}}{3}\right)
    \\
    \times
    \left[
    f_\nu\left(\frac{m_\nu}{T_{\nu,0}(1+z)}\right)+2
    \right]\,
    \Omega_\gamma\,(1+z)^4 ,
\end{multline}
adapted to the case of one neutrino with mass $m_\nu$ and two massless neutrinos, as in Ref.~\cite{Bansal:2025ipo}. Here, $T_{\nu,0}=(4/11)^{1/3}T_{{\rm CMB},0}$ is the present-day neutrino temperature, and $f_\nu$ is a fitting function that interpolates between the massive neutrino's relativistic behavior at early times and non-relativistic behavior at late times. We have verified that this approximation matches the predictions of \texttt{CAMB}~\cite{Lewis:1999bs,Howlett:2012mh} to a precision well below that required for the background observables considered in this work.

\subsubsection{Sound Horizon}

To calculate the sound horizon at the baryon drag epoch, we use the approximation from Ref.~\cite{Brieden:2022heh},
\begin{multline}
    r_{\rm d}\simeq 147.05\,{\rm Mpc}\\
    \times\left(\frac{\Omega_{\rm b} h^2}{0.02236}\right)
    \left(\frac{\Omega_{\rm bc} h^2}{0.1432}\right)^{-0.23}
    \left(\frac{N_{\rm eff}}{3.04}\right)^{-0.1}\,.
    \label{eq:rdrag}
\end{multline}
Here, $h=H_0/(100\,{\rm km\,s}^{-1}{\rm Mpc}^{-1})$ is the dimensionless Hubble constant, $N_{\rm eff}$ is the effective number of relativistic species, and $\Omega_{\rm bc}h^2$ is the physical baryon and cold dark matter density. We have verified that this approximation matches \texttt{CAMB} to a precision below that required by current BAO observations. The impact of the recombination history assumed by Eq.~\eqref{eq:rdrag} on our reconstructions is assessed in Sec.~\ref{sec:extensions-robustness}, where $r_{\rm d}$ is promoted to a free parameter.

Calculating the CMB acoustic scale requires predictions for the recombination redshift, $z_*$, and the corresponding sound horizon, $r_*$. Although approximations exist for these quantities~\cite{Eisenstein:1997ik}, we found they lead to $\theta_*$ values that differ from the outputs of \texttt{RECFAST}~\cite{Seager:1999bc,Wong:2007ym} within \texttt{CAMB} by more than the observational uncertainty on $\theta_*$. For this reason, we calculate $r_*$ and $z_*$ directly using \texttt{RECFAST}.

\subsubsection{\label{sec:fiducial-parameters}Fiducial Parameters}

We assume the following fiducial parameters: a spatially flat Universe, $\Omega_k=0$, the approximate minimum normal-hierarchy neutrino mass sum allowed by oscillation experiments, $\sum m_\nu=0.06\,{\rm eV}$~\cite{ParticleDataGroup:2024cfk}, the effective number of relativistic species predicted by the standard model, $N_{\rm eff}=3.044$~\cite{Froustey:2020mcq}, the present-day CMB temperature measured by \texttt{COBE-FIRAS}, $T_{\rm CMB,0}=2.7255\,{\rm K}$~\cite{Fixsen:2009ug}, and the primordial helium fraction from V1.1 of the \texttt{PArthENoPE} BBN code, $Y_{\rm H}=0.2454$~\cite{Pisanti:2007hk}. The effects of these parameter choices are assessed in Sec.~\ref{sec:extensions-robustness}.

\subsubsection{Cosmological Assumptions}

In contrast to the fiducial parameter assumptions described above, whose effects can be quantified by allowing these parameters to vary freely, we make other cosmological assumptions whose effects are less easily determined. In all of our reconstructions, we assume (\textit{i}) a background Friedmann metric for redshifts $z\gtrsim0.01$ and (\textit{ii}) the validity of the Einstein equations on these scales. When reconstructing $w_{\rm DE}(z)$ or discussing the implications of the $f_{\rm DE}(z)$ reconstructions for the equation of state, we also assume (\textit{iii}) that dark energy behaves like a perfect fluid minimally coupled to matter.

Our analysis is particularly reliant on the first assumption, whose validity remains a topic of debate~\cite{Aluri:2022hzs}. The second and third assumptions are deceptively general in that models violating these conditions often still produce effects that can be captured by the phenomenological reconstructions performed here. As an example, modified gravity models with non-minimal interactions between dark energy and the Ricci scalar define an effective dark energy equation of state capable of crossing the phantom divide, $w_{\rm DE}=-1$~\cite{Perivolaropoulos:2005yv,Chudaykin:2024gol,Ye:2024ywg}, which can be identified by both $f_{\rm DE}(z)$ and $w_{\rm DE}(z)$ reconstructions (Sec.~\ref{sec:de-reconstruction}). However, observing such effects in our reconstructions does not provide any guidance on the underlying physical cause; returning to the example above, a phantom crossing may also signify that dark energy is composed of multiple interacting scalar fields~\cite{Feng:2004ad,Wei:2005nw,Caldwell:2005ai}. Our analysis, incorporating data from the background geometry of the Universe alone, can lend support to such scenarios but is currently incapable of distinguishing between them.

\subsection{\label{sec:de-reconstruction}Dark Energy Reconstruction}

Dark energy contributes to the expansion history through the density parameter $\Omega_{\rm DE}(z) = \Omega_{\rm DE}\,f_{\rm DE}(z)$, where $f_{\rm DE}(z) \equiv \rho_{\rm DE}(z)/\rho_{\rm DE,0}$ is the normalized energy density. For a separately conserved dark energy component, the evolution of $f_{\rm DE}(z)$ is given by
\begin{equation}
    f_{\rm DE}(z) = \exp \left[ 3 \int_{0}^{z} \frac{1 + w_{\rm DE}(z')}{1 + z'} \, dz' \right]\,.
    \label{eq:fde}
\end{equation}
Here, $w_{\rm DE}(z)\equiv p_{\rm DE}/\rho_{\rm DE}$ is the dark energy equation of state, defined as the ratio of its pressure to its energy density. By taking the derivative of Eq.~\eqref{eq:fde},
\begin{equation}
    \frac{\mathrm{d}f_{\rm DE}(z)}{\mathrm{d}z}=3f_{\rm DE}(z)\,\frac{1+w_{\rm DE}(z)}{1+z}\,,
    \label{eq:fde-deriv}
\end{equation}
we see that a \textit{phantom crossing}, where the equation of state passes through $w_{\rm DE}=-1$, occurs when $f_{\rm DE}(z)$ has a local extremum~\cite{DiValentino:2020naf,Adil:2023exv,Caldwell:2025inn}.

When reconstructing dark energy, one must decide which function to constrain directly, with the two principal choices being $f_{\rm DE}(z)$ and $w_{\rm DE}(z)$. While the equation of state is more closely connected to the microphysics of dark energy~\cite{Copeland:2006wr}, Eq.~\eqref{eq:fde} shows that its redshift evolution is more difficult to constrain using background data: barring direct probes of the Hubble parameter, distance measurements at redshift $z$ depend on $f_{\rm DE}(z')$ for all $z'<z$, which itself depends on $w_{\rm DE}(z'')$ for all $z''<z'$. Energy density amplitudes at different redshifts are therefore less mutually degenerate when predicting these observables, leading to tighter constraints. Although a reconstructed $f_{\rm DE}(z)$ can in principle be mapped to $w_{\rm DE}(z)$ through Eq.~\eqref{eq:fde-deriv}, the result depends on the assumed functional form of $f_{\rm DE}(z)$ and becomes ill-defined when $\mathrm{d}f_{\rm DE}(z)/\mathrm{d}z=0$, as occurs in piecewise-constant reconstructions. For these reasons, we reconstruct both functions directly.

The next decision is the reconstruction methodology, \textit{i.e.}, the parametric form chosen for the reconstructed function. Let us denote this function by $g(z)$. There are three main approaches: (\textit{i}) expanding $g(z)$ in basis functions, (\textit{ii}) parameterizing $g(z)$ by its amplitudes at particular redshifts, interpolated by a spline, and (\textit{iii}) representing $g(z)$ as a piecewise-constant function of redshift, sometimes known as \textit{binning}. The first two approaches involve additional assumptions: the basis functions, the order of the expansion, the number and location of the spline points, and the interpolation scheme, optionally including a smoothness prior that prevents wild oscillations. These choices can introduce correlations between the reconstruction parameters that tighten constraints at the expense of potential biases. The advantage of the third methodology is that no additional correlations are introduced \textit{a priori}, apart from those inherent to the observables, and the only freedom lies in choosing the total number of bins and their redshift ranges. Since our goal is to minimize assumptions and remain as data-driven as possible, we adopt the piecewise-constant approach.

Hence, we represent $g(z)\in\left\{f_{\rm DE}(z),\,w_{\rm DE}(z)\right\}$ as
\begin{equation}
    g(z)=
    \begin{cases}
        g_0 & z\in[z_0,z_1)\,,\\
        g_1 & z\in[z_1,z_2)\,,\\
        \vdots & \vdots \\
        g_{n-1} & z\in[z_{n-1},z_n)\,,\\
        g_{\Lambda} & z\geq z_n\,,
    \end{cases}
    \label{eq:binned-function}
\end{equation}
where $\left\{g_i\right\}_{i=0}^{n-1}$ are the reconstructed amplitudes and $\left\{z_i\right\}_{i=0}^{n}$ are the boundaries of their redshift ranges. Requiring that the reconstructed function approaches its cosmological constant limit, $g_{\Lambda}$, deep into the matter domination epoch is standard practice in dark energy reconstructions, as dark energy is subdominant during these times and is therefore difficult to constrain. Replacing $g_{\Lambda}$ with a free parameter amounts to removing the CMB prior on $\theta_*$, the only data point probing $z>z_n$, and this extension is tested in Sec.~\ref{sec:extensions-robustness}.

In traditional BAO analyses, galaxy populations spanning largely non-overlapping redshift ranges are used to infer Alcock-Paczynski distortion measurements, which then determine cosmological distances at a corresponding number of effective redshifts~\cite{DESI:2024uvr,Chen:2024tfp}. We follow Ref.~\cite{Bansal:2025ipo} in aligning our reconstruction bins with the redshift ranges of the DESI-BAO tracers, as listed in Table~III of Ref.~\cite{DESI:2025zgx}. This choice has several desirable features. First, it ensures that each bin contains at least one BAO measurement, providing more direct constraints through $D_{\rm H}/r_{\rm d}$ or its contribution to $D_{\rm V}/r_{\rm d}$. Furthermore, the effective redshift of each BAO measurement has a theoretical uncertainty equal to the redshift range of the corresponding tracer~\cite{Chen:2024tfp}. Our binning choice is the only one that is fully robust to this uncertainty.\footnote{This is only approximately true for our reconstructions using BAO from SDSS, whose tracer redshift ranges differ slightly from those of DESI (Table~\ref{tab:bao-sne}).} By including SNe~Ia data, the number of reconstruction parameters remains much smaller than the total number of distance measurements, mitigating the risk of overfitting. In addition to the DESI-motivated bins, we include one additional bin at low redshifts, $z\in[0,0.1)$, where there are no BAO measurements but many SNe~Ia observations (Sec.~\ref{sec:datasets}).

There are some caveats to the binning choice for $f_{\rm DE}(z)$. First, one must remain consistent with the normalization condition $f_{\rm DE}(0)=1$, which follows directly from the definition of this function. In the DESI reconstructions~\cite{DESI:2025fii}, this was achieved by fixing $f_{\rm DE}(z)$ to unity over the small interval $z\in[0,z_{\rm min})$ and using $z_{\rm min}\sim0.05$ as the minimum redshift of the first reconstruction amplitude~\cite{Matthewson:2026}. We adopt this strategy here, fixing $f_{\rm DE}(z)=1$ within $z\in[0,z_{\rm min})$ and restricting the first amplitude to $[z_{\rm min},0.1)$.

Ideally, one would choose $z_{\rm min}\sim0.01$ so that no SNe~Ia measurements are omitted from the reconstruction. The issue is that binned $f_{\rm DE}(z)$ reconstructions suffer from a degeneracy between the first amplitude and the Hubble constant. In our reconstructions, leaving this degeneracy unconstrained tends to increase the marginalized $f_{\rm DE}(z)$ constraints while decreasing $H_0$, and this pull towards larger $f_{\rm DE}(z)$ values becomes stronger as $z_{\rm min}$ approaches zero, leading to biased constraints. Our solution is to choose $z_{\rm min}$ for each BAO and SNe~Ia combination such that the reconstructions show no bias when applied to mock data generated from $\Lambda$CDM and CPL cosmologies. We confirmed that our $z_{\rm min}$ choices are suitable for a wide range of CPL cosmologies, including those with extreme evolution at low redshifts. Our mock data tests are explained in more detail in Appendix~\ref{sec:mocks}, and our binning choices are summarized in Table~\ref{tab:binning}.

\begin{table}
    \setlength{\tabcolsep}{2pt}
    \footnotesize
    \resizebox{\columnwidth}{!}{
    \begin{tabular}{lcc}
        \toprule
        Function & {$\left\{z_i\right\}_{i=0}^n$} & {When} \\
        \midrule
        $f_{\rm DE}(z)$ & $\left\{0,z_{\rm min}, 0.1, 0.4, 0.6, 0.8, 1.1, 1.6, 4.2\right\}$ & \\
        & $z_{\rm min}=0.04$ & Union3.1 Incl. \\
        & $z_{\rm min}=0.02$ & Pantheon+ Incl. \\
        & $z_{\rm min}=0.03$ & DES-Dovekie Incl. \\
        \midrule
        $w_{\rm DE}(z)$ & $\left\{0, 0.1, 0.4, 0.6, 0.8, 1.1, 1.6, 4.2\right\}$ & Always \\
        \bottomrule
    \end{tabular}
    }
    \caption{
        Binning choices used to reconstruct dark energy in the piecewise-constant representation of Eq.~\eqref{eq:binned-function}. The equation of state is always reconstructed in the DESI-motivated bins, with boundaries $\{0.1,\ldots,4.2\}$, together with an additional low-redshift bin, $z\in[0,0.1)$. By contrast, the normalized energy density must satisfy $f_{\rm DE}(0)=1$ by definition, so it is fixed to unity over $z\in[0,z_{\rm min})$. The choice of $z_{\rm min}$ affects the accuracy of the marginalized $f_{\rm DE}(z)$ posteriors, which is determined from performance on mock datasets, as discussed in Appendix~\ref{sec:mocks}.
    }
    \label{tab:binning}
\end{table}

\subsection{Parameter Inference}

\begin{table}
    \setlength{\tabcolsep}{4pt}
    \footnotesize
    \begin{tabular}{lcc}
        \toprule
        {Parameter} & {Prior} & {Reference} \\
        \midrule
        \multicolumn{3}{l}{\textit{Uniform Priors}} \\
        \midrule
        $\Omega_{\rm c} h^2$
            & $\mathcal{U}(0.001,\, 0.99)$
            & --- \\
        $f_{\rm DE}(z)$
            & $\mathcal{U}(-1,\, 3)$
            & --- \\
        $w_{\rm DE}(z)$
            & $\mathcal{U}(-3,\, 1)$
            & --- \\
        \midrule
        \multicolumn{3}{l}{\textit{Physical (Gaussian) Priors}} \\
        \midrule
        $\Omega_{\rm b} h^2$
            & $\mathcal{N}(0.02218,\, 0.00055^2)$
            &~\cite{Schoneberg:2024ifp, Burns:2023sgx} \\
        $100\theta_*$
            & $\mathcal{N}(1.04110,\, 0.00053^2)$
            &~\cite{DESI:2025zgx,Planck:2018vyg} \\
        \bottomrule
    \end{tabular}
    \caption{
        Priors on our cosmological and reconstruction parameters. The priors on the physical cold dark matter density, the physical baryon density, and the angular sound horizon at recombination follow the choices of DESI~\cite{DESI:2025zgx,DESI:2025fii}. In particular, $\Omega_{\rm b}h^2$ is sampled from the Gaussian prior derived in Ref.~\cite{Schoneberg:2024ifp}, while $\theta_*$ is sampled from its \textit{Planck}~2018 posterior under $\Lambda$CDM~\cite{Planck:2018vyg}, with the uncertainty increased by $\sim 75\%$ to accommodate shifts in alternative models~\cite{DESI:2025zgx}. The priors on the dark energy bins are symmetric around their $\Lambda$CDM limits, $f_{\rm DE}(z)=1$ and $w_{\rm DE}(z)=-1$.
    }
    \label{tab:priors}
\end{table}

In addition to the seven reconstruction parameters for $f_{\rm DE}(z)$ or $w_{\rm DE}(z)$ and the fiducial constants in Sec.~\ref{sec:fiducial-parameters}, three additional cosmological parameters are needed to predict the BAO and SNe~Ia observables. These are the physical cold dark matter density, $\Omega_c h^2$, the physical baryon density, $\Omega_b h^2$, and the angular sound horizon at recombination, $\theta_*$.\footnote{We use physical energy densities, $\sim\Omega h^2$, and the angular sound horizon, $\theta_*$, to match the conventions of DESI~\cite{DESI:2025fii}, although energy densities, $\sim\Omega$, and the Hubble constant, $H_0$, can be used with equivalent results. In our case, $H_0$ is derived from $\theta_*$ and the other free parameters using a root-finding algorithm, and the $\Omega$ parameters are then derived from the physical densities using the resulting value of $H_0$.} To estimate the posteriors of these parameters from the data, we use the Markov Chain Monte Carlo (MCMC) sampler in \texttt{Cobaya}~\cite{Torrado:2020dgo} together with our custom theory module. To obtain Bayesian evidences, we use the nested sampler \texttt{PolyChord}~\cite{Handley:2015fda,Handley:2015vkr}. In each case, we employ the priors given in Table~\ref{tab:priors}. These priors are mostly flat, apart from a Gaussian prior on $\Omega_b h^2$ from Big Bang Nucleosynthesis (BBN), calculated in Ref.~\cite{Schoneberg:2024ifp} using the \texttt{PRyMordial} code~\cite{Burns:2023sgx}, and a conservative Gaussian prior on $\theta_*$ from \textit{Planck} with a $75\%$ enlarged uncertainty to accommodate shifts in alternative models, matching the analysis choices of DESI~\cite{DESI:2025zgx}.

We use \texttt{GetDist}~\cite{Lewis:2019xzd} and \texttt{anesthetic}~\cite{Handley:2019mfs} to analyze the results of our MCMC and nested sampling runs, respectively.

\subsection{\label{sec:statistics}Statistics}

To concisely analyze the results of the reconstructions, we use summary statistics quantifying their agreement with fiducial expectations and their internal consistency. Of particular interest is the statistical agreement between the reconstructions and the null hypothesis that the Universe is described by a $\Lambda$CDM cosmology. We first assess this agreement locally, as a function of redshift, by considering the significance of deviations from the $\Lambda$CDM predictions $f_{\rm DE}(z)=1$ and $w_{\rm DE}(z)=-1$ in each redshift bin. By calculating the $p$-values of these predictions with respect to the bins' posteriors and converting them to equivalent Gaussian $\sigma$ values using the standard normal distribution, we obtain an intuitive tomographic view of the consistency of $\Lambda$CDM with current expansion history measurements.

Because these local significances are evaluated across several redshift bins, dataset combinations, and reconstructed functions, they may be subject to the look-elsewhere effect, whereby significant `discoveries' might occur in a large number of trials due to random chance. Ref.~\cite{Ong:2025cwv} accounted for this effect by considering the threshold, $\sigma_{\rm thresh}\equiv\sqrt{2}\mathrm{Erfc}^{-1}(1/N)$, above which one expects minimal false positives in $N$ independent trials. As a first estimate, we can take
\begin{equation*}
    N\sim(7\text{ bins})\times(6\text{ BAO-SNe~Ia combinations})=42 \text{ trials}
\end{equation*}
to obtain the threshold $\sigma_{\rm thresh}\sim2.3$ for each of $f_{\rm DE}(z)$ and $w_{\rm DE}(z)$. However, this simple estimate neglects the fact that our `trials' are highly correlated; there are therefore many fewer \textit{independent} trials, which should face a less severe look-elsewhere threshold. A formal calculation would require determining the effective number of independent tests, accounting for the correlations between bins and between dataset combinations. We defer this calculation to future work. Here, we use the local significances to identify the redshift ranges in which the data prefer departures from $\Lambda$CDM, while refraining from making strong statements about their true corrected significance.

Global comparisons between the reconstructions and $\Lambda$CDM are possible through the improvement in the minimum chi-square, $\Delta\chi^2\equiv\chi^2_{\rm recon}-\chi^2_{\Lambda}$, which can be converted into an equivalent Gaussian significance using Wilks' theorem for nested models~\cite{Wilks:1938dza}, and the difference in Bayesian evidence, $\Delta\log B\equiv\log B_{\rm recon}-\log B_{\Lambda}$. Although we consider these global statistics in Sec.~\ref{sec:model-comparison}, one should note that flexible reconstructions are not necessarily expected to perform well according to these metrics compared to minimal physical models. Whereas these statistics penalize models for introducing more freedom than is required to fit the data, reconstructions deliberately introduce many degrees of freedom to explore the data's preferences across redshift. Furthermore, the Bayesian evidence depends on our priors on $f_{\rm DE}(z)$ and $w_{\rm DE}(z)$, which do not have strong physical motivation, and this statistic is subject to large fluctuations under physically equivalent reparametrizations (e.g., sampling $h$ rather than $\theta_*$). These can be viewed as desirable features only when one has careful and explicit motivations for the prior choices.\footnote{\label{footnote:bayes-factor}Bayesian model comparison is possibly at odds with the philosophy of reconstructions more generally. The Bayesian evidence is penalized by the ratio of posterior to prior probability, whereas agnostic reconstructions have intentionally wide priors to explore a range of possible behaviors. This leads to our $f_{\rm DE}(z)$ reconstructions having lower evidence than our $w_{\rm DE}(z)$ reconstructions simply because the former has a more constrained posterior but the same prior volume.}

The internal consistency of the reconstructions can be quantified using the suspiciousness statistic, $S$~\cite{Handley:2019wlz}. This statistic builds on the Bayesian evidence ratio, $R=B_{12}/(B_1B_2)$, where $B_{12}$ is the evidence obtained from the combined datasets 1 and 2, while $B_1$ and $B_2$ are those obtained from these datasets analyzed separately. An $R>1$ supports the hypothesis that a common set of parameters can describe the datasets, signaling mutual consistency~\cite{Marshall:2004zd}. The suspiciousness statistic attempts to remove the prior dependence in $R$ by subtracting the prior-dependent term arising from perfectly Gaussian posteriors. For such posteriors, $S$ is chi-squared distributed and can be mapped to an equivalent Gaussian significance that summarizes the agreement between the combined datasets under the assumed model. As our posteriors are approximately Gaussian (Figs.~\ref{fig:fde-triangle} and \ref{fig:wde-triangle}), we compute such significances and report them in Table~\ref{tab:bayesian-stats}.

\subsection{Principal Components}

Constraints on $f_{\rm DE}(z)$ and $w_{\rm DE}(z)$ at different redshifts are correlated through the integral relations connecting dark energy to cosmological distances (Sec.~\ref{sec:background}). These correlations lead to parameter degeneracies that broaden the marginalized constraints and can obscure how many independent departures from $\Lambda$CDM the data prefer. For these reasons, it is useful to consider linear combinations of the reconstructed amplitudes that yield uncorrelated estimates of $f_{\rm DE}(z)$ and $w_{\rm DE}(z)$ at different redshifts. Such estimates are provided by \textit{principal component analysis} (PCA).

We reconstruct $f_{\rm DE}(z)$ and $w_{\rm DE}(z)$ as piecewise-constant functions of redshift. This representation was given in Eq.~\eqref{eq:binned-function} for $g(z)\in\left\{f_{\rm DE}(z),\,w_{\rm DE}(z)\right\}$ and can be rewritten as
\begin{equation}
    g(z)=\sum_{i=0}^{n-1}g_i\,e_i(z)\,.
    \label{eq:g-sum}
\end{equation}
Here, the basis functions $e_i(z)$ are top-hat functions spanning $z\in[z_i,z_{i+1})$, and their coefficients $g_i$ are the reconstructed amplitudes. The covariance matrix, $C_g$, of these amplitudes is defined as
\begin{equation}
    C_g=\langle \bm g\bm g^T\rangle
    - \langle\bm g\rangle \langle\bm g^T\rangle\,.
    \label{eq:g-covariance}
\end{equation}
We seek a transformation, or weight matrix, $W$, that defines new amplitudes $\hat{\bm g}=W\bm g$ with diagonal covariance, $C_{\hat g}$. Substituting $\hat{\bm g}$ for $\bm g$ in Eq.~\eqref{eq:g-covariance} gives
\begin{equation}
    C_{\hat g} = W C_g W^T\,.
    \label{eq:ghat-covariance}
\end{equation}
One suitable transformation is obtained by diagonalizing the inverse covariance matrix of the original amplitudes,
\begin{equation}
    C_g^{-1}=V^TD V\,,
\end{equation}
where $V$ is the orthogonal matrix whose rows are the eigenvectors of $C_g^{-1}$, and $D$ is the diagonal matrix of eigenvalues. Choosing $W=V$ gives uncorrelated amplitudes with covariance $C_{\hat g}=D^{-1}$, as follows from Eq.~\eqref{eq:ghat-covariance}. This is the traditional form of PCA, first applied to dark energy reconstructions in Ref.~\cite{Huterer:2002hy}. Since $V$ is orthogonal, Eq.~\eqref{eq:g-sum} shows that $\hat{\bm g}$ can be interpreted as the amplitudes of the new basis functions, $\hat{\bm e}=W\bm e$. 

In traditional PCA, the amplitudes $\hat g_i$ are ordered by increasing variance, $\mathrm{var}(\hat g_i)$. The corresponding basis functions $\hat{\bm e}$ generally extend over the full redshift range and oscillate around zero, making them difficult to associate with localized redshift intervals~\cite{Huterer:2002hy}. While traditional PCA provides an optimal linear compression of the original reconstruction~\cite{Kessy_2018}, the delocalization of its basis functions complicates physical interpretation.

An alternative form of decorrelation---sometimes called localized principal component analysis (LPCA) in the cosmology literature~\cite{Huterer:2004ch,Hamilton:1999uw} or zero-phase component analysis (ZCA) more broadly---produces uncorrelated amplitudes that remain as close as possible to the original amplitudes and, therefore, basis functions that are optimally localized around the original basis~\cite{Kessy_2018}. In the context of binned reconstructions, this gives amplitudes $\hat g_i$ whose deviations from $\Lambda$CDM can be more clearly associated with compact redshift intervals centered around the original top-hat bins. This property motivated the original applications of LPCA in cosmology~\cite{Hamilton:1999uw,Huterer:2004ch} and makes this decorrelation scheme ideal for our purposes.

The unnormalized LPCA weights are defined by the positive square root of $C_g^{-1}$,
\begin{equation}
    W=C_g^{-1/2}=V^TD^{1/2}V\,,
\end{equation}
which yields uncorrelated amplitudes with unit covariance, $C_{\hat g}=I_n$. In the context of functional reconstructions, it is useful to normalize the rows of $W$ so that they sum to unity,
\begin{multline}
    W\rightarrow N\cdot W\,,\\\quad N=\mathrm{diag}\left\{\frac{1}{\sum_{j=0}^{n-1} W_{ij}}:i=0,\dots,n-1\right\}\,.
    \label{eq:w-norm}
\end{multline}
This normalization ensures that the LPCA amplitudes,
\begin{equation}
    \hat g_i=\sum_{j=0}^{n-1} W_{ij}\,g_j\,,
\end{equation}
have the same physical units as the original amplitudes and can reliably identify deviations from constancy.\footnote{To see this, assume that the true underlying function is constant, $g(z)=g_{\rm true}$, and that the original amplitudes recover this value, $\langle g_j\rangle=g_{\rm true}$ for all $j$. Then, $\langle \hat g_i\rangle=\sum_{j}W_{ij}\langle g_j\rangle =g_{\rm true}\sum_{j}W_{ij}$, so the transformed amplitudes also recover $g_{\rm true}$ only if each row of $W$ sums to unity.} This row normalization distinguishes LPCA from the standard ZCA transformation, resulting in weights that do not strictly maximize the similarity between the original and transformed amplitudes. After normalization, the transformed amplitudes remain uncorrelated, but their covariance is no longer the identity; instead, $C_{\hat g}=N N^T$, with $N$ defined in Eq.~\eqref{eq:w-norm}.

\section{\label{sec:datasets}Datasets}

\begin{table}
    \resizebox{\columnwidth}{!}{
    \begin{tabular}{lllcccc}
    \toprule
    {Collaboration} & {Population} & {$z$ Range} & {$z_\mathrm{eff}$} & {Observable} & {$N_{\rm data}$} & {Ref.} \\
    \midrule
    \multicolumn{7}{l}{\textit{Baryon Acoustic Oscillations (BAO)}} \\
    \midrule
    \multirow{8}{*}{\parbox{2.1cm}{SDSS}}
        & MGS
        & $[0.07,\, 0.20]$
        & $0.15$
        & $D_{\rm V}/r_{\rm d}$
        & $1$
        & \multirow{8}{*}{\parbox{0cm}{\cite{eBOSS:2020fvk}}} \\
        & BOSS
        & $[0.2,\, 0.5]$
        & $0.38$
        & $(D_{\rm M}/r_{\rm d},D_{\rm H}/r_{\rm d})$
        & $2$
        & \\
        & BOSS
        & $[0.4,\, 0.6]$
        & $0.51$
        & $(D_{\rm M}/r_{\rm d},D_{\rm H}/r_{\rm d})$
        & $2$
        & \\
        & LRG
        & $[0.6,\,1.0]$
        & $0.70$
        & $(D_{\rm M}/r_{\rm d},D_{\rm H}/r_{\rm d})$
        & $2$
        & \\
        & ELG
        & $[0.6,\,1.1]$
        & $0.85$
        & $D_{\rm V}/r_{\rm d}$
        & $1$
        & \\
        & QSO
        & $[0.8,\,2.2]$
        & $1.48$
        & $(D_{\rm M}/r_{\rm d},D_{\rm H}/r_{\rm d})$
        & $2$
        & \\
        & $\text{Ly}\alpha$-$\text{Ly}\alpha$
        & $z>2.1$
        & $2.33$
        & $(D_{\rm M}/r_{\rm d},D_{\rm H}/r_{\rm d})$
        & $2$
        & \\
        & $\text{Ly}\alpha$-QSO
        & $z>1.77$
        & $2.33$
        & $(D_{\rm M}/r_{\rm d},D_{\rm H}/r_{\rm d})$
        & $2$
        & \\
    \midrule
    \multirow{7}{*}{\parbox{2.1cm}{DESI-DR2}}
        & BGS
        & $[0.1,\,0.4]$
        & $0.30$
        & $D_{\rm V}/r_{\rm d}$
        & $1$
        & \multirow{7}{*}{\parbox{0cm}{\cite{DESI:2025zgx}}} \\
        & LRG1
        & $[0.4,\,0.6]$
        & $0.51$
        & $(D_{\rm M}/r_{\rm d},D_{\rm H}/r_{\rm d})$
        & $2$
        & \\
        & LRG2
        & $[0.6,\,0.8]$
        & $0.71$
        & $(D_{\rm M}/r_{\rm d},D_{\rm H}/r_{\rm d})$
        & $2$
        & \\
        & $\text{LRG3}+\text{ELG1}$
        & $[0.8,\,1.1]$
        & $0.93$
        & $(D_{\rm M}/r_{\rm d}\,,D_{\rm H}/r_{\rm d})$
        & $2$
        & \\
        & ELG2
        & $[1.1,\,1.6]$
        & $1.32$
        & $(D_{\rm M}/r_{\rm d},D_{\rm H}/r_{\rm d})$
        & $2$
        & \\
        & QSO
        & $[0.8,\,2.1]$
        & $1.48$
        & $(D_{\rm M}/r_{\rm d},D_{\rm H}/r_{\rm d})$
        & $2$
        & \\
        & Ly$\alpha$
        & $[1.8,\,4.2]$
        & $2.33$
        & $(D_{\rm M}/r_{\rm d},D_{\rm H}/r_{\rm d})$
        & $2$
        & \\
    \midrule
    \multicolumn{7}{l}{\textit{Type~Ia Supernovae (SNe~Ia)}} \\
    \midrule
        Union3.1
        & ---
        & $[0.05,\,2.26]$
        & ---
        & $\mu$
        & $22$
        &~\cite{Hoyt:2026fve} \\
        Pantheon$+$
        & ---
        & $(0.01,\,2.26]$
        & ---
        & $\mu$
        & $1588$
        &~\cite{Brout:2022vxf} \\
        DES-Dovekie
        & ---
        & $[0.025,\,1.3]$
        & ---
        & $\mu$
        & $1820$
        &~\cite{Popovic:2025glk, DES:2025sig} \\
    \bottomrule
    \end{tabular}
    }
    \caption{
        Selected characteristics of the BAO and SNe~Ia datasets used in this paper. For each dataset, we give the redshift range, the constrained observable, and the number of data points. For SNe~Ia, the direct observable is the apparent magnitude, but constraints on the distance modulus, $\mu$, are obtained by marginalizing over the absolute magnitude during parameter estimation. For the BAO datasets, measurements are subdivided by galaxy population, or, in some cases, by survey, and we list the effective redshift, $z_{\rm eff}$, at which each observable is reported. The data points in Union3.1 result from compressing $\sim2000$ SNe~Ia observations~\cite{Hoyt:2026fve}.
    }
    \label{tab:bao-sne}
\end{table}

\begin{figure}
    \centering
    \includegraphics[width=\linewidth]{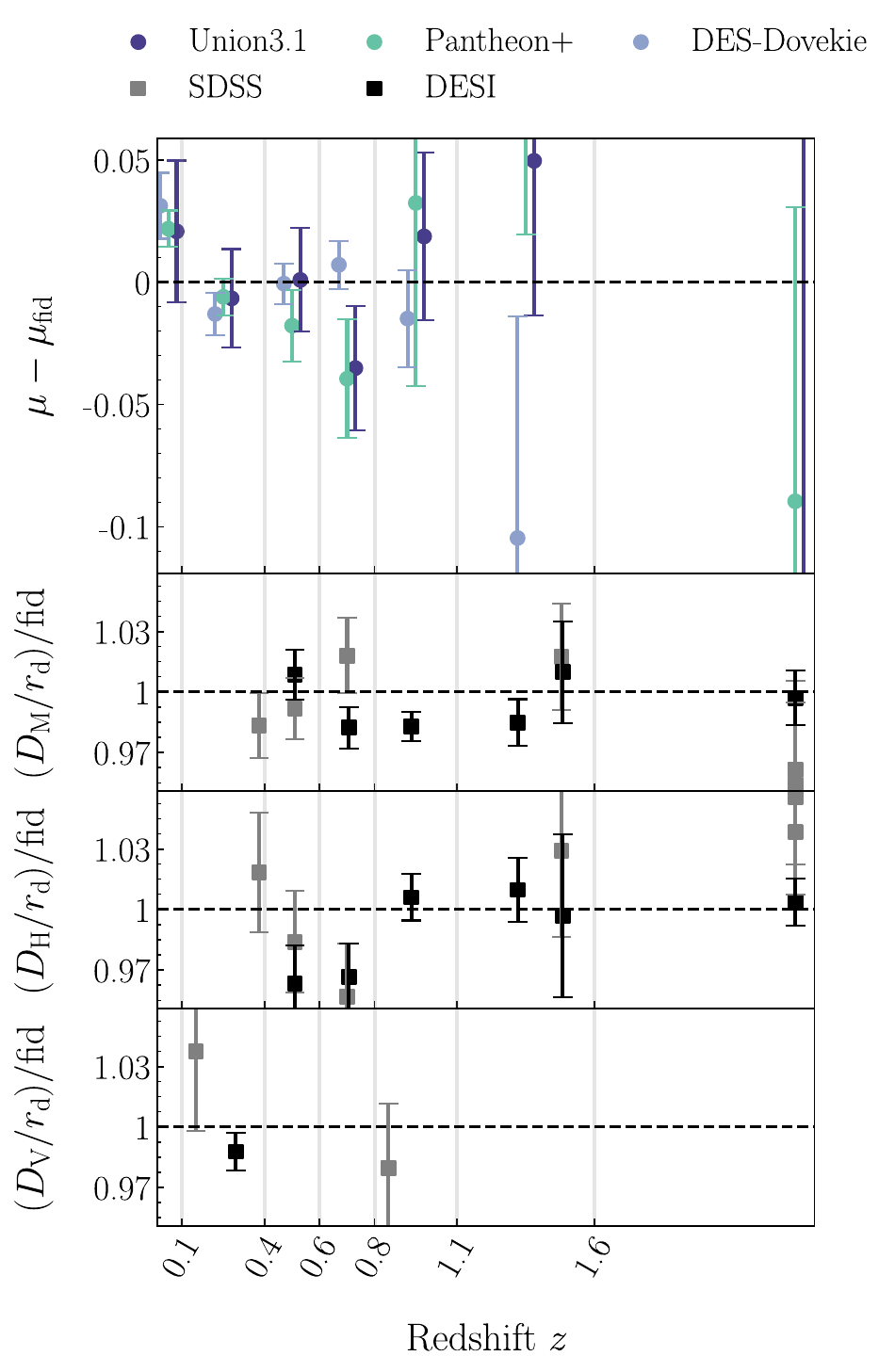}
    \caption{Residuals of the BAO and SNe~Ia observations with respect to the fiducial \textit{Planck} $\Lambda$CDM model, $(\Omega_ch^2,\,\Omega_bh^2,\,h)\simeq(0.120,0.0224,0.674)$~\cite{Planck:2018vyg}. For visual purposes only, we compress the SNe~Ia observations to their weighted averages~\cite{Schmelling:1994pz} within the redshift bins of the reconstruction (after subtracting the weighted average to correct for differences in absolute magnitude calibration, as in Ref.~\cite{DESI:2025zgx}), indicated by the $x$-axis ticks and gray vertical lines. The directions of the residuals in the top two rows can be directly compared, since $\mu-\mu_{\rm fid}\propto \log(D_{\rm M}/D_{{\rm M},{\rm fid}})$ at fixed redshift.}
    \label{fig:bao-sne}
\end{figure}

\begin{figure}
    \centering
    \includegraphics[width=\linewidth]{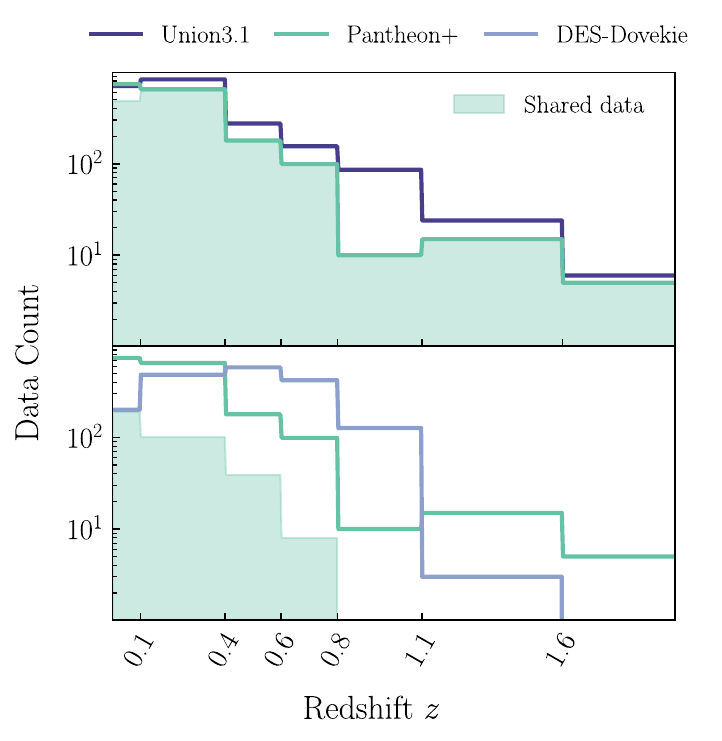}
    \caption{
        Number of SNe~Ia observations in each redshift bin of the reconstructions. The solid lines show the number counts for each catalog; Pantheon+ is shown in both the upper and lower panels for comparison with Union3.1 and DES-Dovekie, respectively. The shaded bands show the number of observations shared by Union3.1 and Pantheon+ in the upper panel, and by DES-Dovekie and Pantheon+ in the lower panel. See the main text and Footnote~\ref{footnote:shared-sne} for further details.}
    \label{fig:shared-sne}
\end{figure}

We employ likelihoods from the most recent and widely used BAO and SNe~Ia catalogs, summarized in Table~\ref{tab:bao-sne}, together with the BBN and conservative CMB acoustic-scale priors given in Table~\ref{tab:priors}. Notably, our baseline reconstructions do not use any additional CMB information, such as an effective constraint on the cold dark matter density inferred within $\Lambda$CDM, which is included in many other analyses. Omitting this information reduces our dependence on assumptions used to infer $\Omega_c h^2$ from the CMB and ensures that our results are not driven by the $\Omega_m$~tension that has been discussed as a possible confounding factor in interpreting indications for evolving dark energy.\footnote{It has been found that assuming $\Lambda$CDM pre-recombination physics while including CMB data in late-time analyses of dark energy can lead to preferences for a phantom crossing~\cite{Ye:2025ark,Shlivko:2026jxa,Wang:2026fuh}. These inclusions fix the angular diameter distance to recombination, $D_{\rm A}(z_*)$, to its value in the $\Lambda$CDM model with the CMB determination of $\Omega_m$; dark energy evolution must then `average' to the constant $\Lambda$CDM limit, leading to one or more phantom crossings. By omitting a CMB prior on $\Omega_m$ in our baseline analysis, we avoid this particular effect.}

As a visual complement to Table~\ref{tab:bao-sne}, Fig.~\ref{fig:bao-sne} shows the residuals of the BAO and SNe~Ia measurements with respect to the predictions of the fiducial \textit{Planck} $\Lambda$CDM model. The top two rows show $\mu$ from SNe~Ia and $D_{\rm M}/r_{\rm d}$ from BAO, and the signs of these residuals can be directly compared, since $\mu-\mu_{\rm fid}\propto \log\left(D_{\rm M}/D_{{\rm M},{\rm fid}}\right)$. The third row shows $D_{\rm H}/r_{\rm d}$ from BAO and provides the most direct visual indication of deviations in the expansion rate relative to $\Lambda$CDM expectations. The minima in $D_{\rm H}/r_{\rm d}$ near $z\sim0.7$ and in the top two rows near $z\sim0.8$ suggest that these datasets prefer a localized enhancement in $f_{\rm DE}(z)$ around the former redshift; see Sec.~\ref{sec:results} for further discussion. The final row shows the BAO datasets' $D_{\rm V}/r_{\rm d}$ residuals. The presence of an SDSS $D_{\rm V}/r_{\rm d}$ measurement at $z\sim0.15$ explains its increased constraining power at low redshifts compared to DESI. Beyond these broad points, this figure will be useful for interpreting the specific features that appear in the reconstructions presented in the following section.

It is worth emphasizing that, although the different SNe~Ia samples were produced by independent collaborations using different methodologies, they share a significant number of observations from previous surveys. These samples are therefore not statistically independent and may be affected by common instrumental or calibration systematics. Fig.~\ref{fig:shared-sne} shows the number of observations shared between these catalogs, Union3.1--Pantheon+ in the top panel and DES-Dovekie--Pantheon+ in the bottom panel, alongside their total numbers of observations.\footnote{\label{footnote:shared-sne}The number of shared observations was determined by matching IDs between the catalogs. We obtained the uncompressed Union observations from the GitHub repository \url{github.com/rubind/union3_release}, and we thank the authors of Ref.~\cite{Hoyt:2026fve} for providing a mapping between the Union3.1 and Pantheon+ IDs.} Overall, we find that around $83\%$ of the observations in Pantheon+ are also included in Union3.1, distributed across the redshift ranges of these datasets, whereas only $\sim20\%$ are contained in DES-Dovekie, mostly at low redshifts. Common preferences between Union3.1 and Pantheon+ are therefore expected to be more statistically correlated than common preferences between either of these datasets and DES-Dovekie.

\section{\label{sec:results}Results}

\subsection{\label{sec:fde-res}Energy Density}

\begin{figure*}
    \centering
    \includegraphics[width=\linewidth]{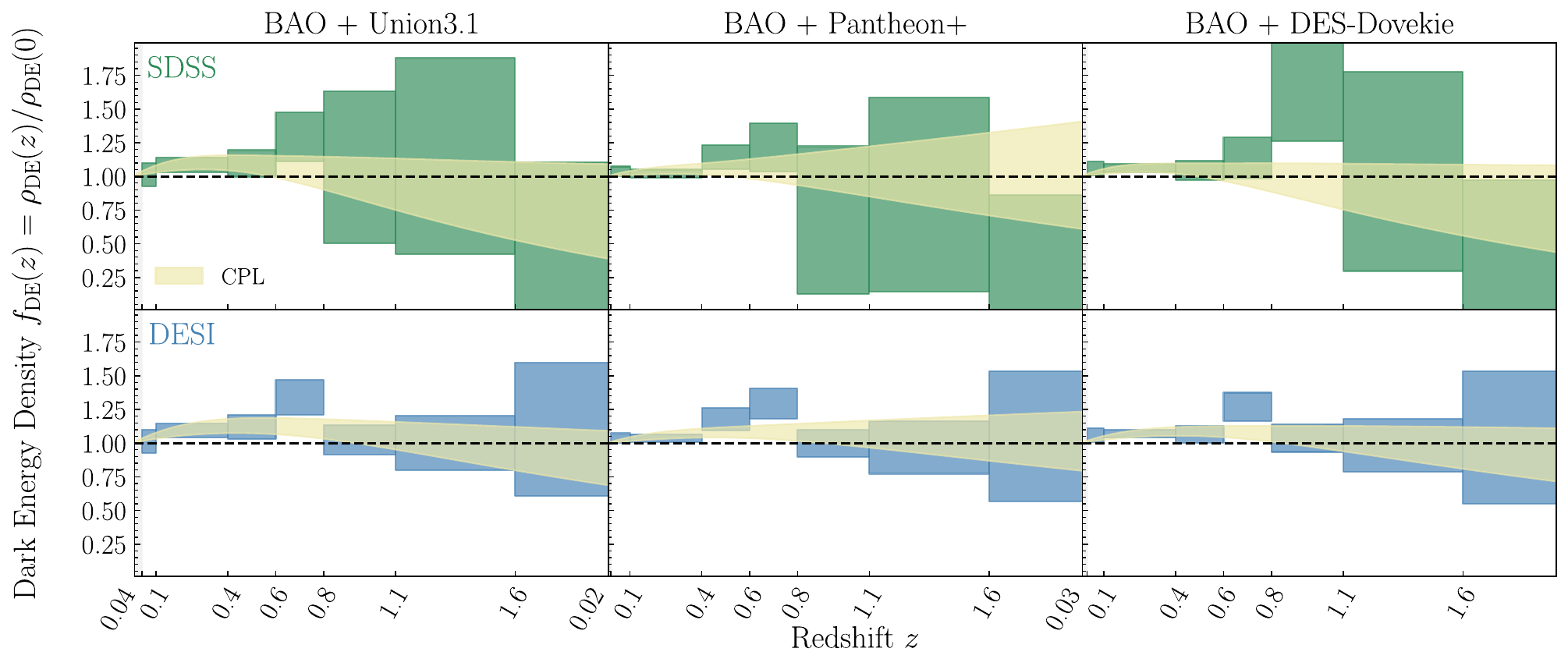}
    \caption[]{Piecewise-constant reconstructions ($1\sigma$ posteriors) of the normalized dark energy density, $f_{\rm DE}(z)\equiv\rho_{\rm DE}(z)/\rho_{\rm DE}(0)$, for each BAO and SNe~Ia combination, calibrated by a BBN prior on $\Omega_bh^2$ and a conservative CMB prior on $\theta_*$ in each case. The top row shows combinations involving BAO from SDSS (green), while the bottom row shows combinations using DESI (blue). The light green bands represent the $1\sigma$ constraints obtained from the CPL parametrization of the dark energy equation of state. The reconstructed $f_{\rm DE}(z)$ is fixed to unity between $0\leq z<z_{\rm min}$ to remain consistent with its definition at $z=0$, where $z_{\rm min}$ depends on the SNe~Ia dataset included (Table~\ref{tab:binning}). Each reconstruction shows a local maximum near $z\sim0.6-0.8$, consistent with a phantom crossing in this redshift range. These maxima are the only notable deviations from a CPL evolution.}
    \label{fig:fde-posteriors}
\end{figure*}

Figure~\ref{fig:fde-posteriors} shows the reconstructed dark energy density, $f_{\rm DE}(z)\equiv \rho_{\rm DE}(z)/\rho_{\rm DE}(0)$, for each BAO and SNe~Ia combination considered. In all cases, the reconstructions are calibrated by a BBN prior on $\Omega_b h^2$ and a conservative CMB prior on $\theta_*$, as described in Sec.~\ref{sec:datasets}.

A remarkable feature of these reconstructions is the qualitative agreement between all six dataset combinations. At low redshifts, within the first reconstructed bin, the dark energy density remains broadly consistent with the $\Lambda$CDM expectation, $f_{\rm DE}=1$. Moving to intermediate redshifts, however, all reconstructions show a rise in $f_{\rm DE}(z)$, reaching an apparent local maximum mostly within the redshift interval $z\in[0.6,0.8)$. The only exception is the SDSS and DES-Dovekie combination, whose apparent maximum is shifted to slightly higher redshifts, $z\in[0.8,1.1)$. Beyond these redshifts, the reconstructions involving DESI return towards the $\Lambda$CDM limit, while those using SDSS fluctuate slightly below this limit in the highest-redshift bin, $z\in[1.6,4.2)$. This difference arises because these datasets disagree slightly on Lyman-$\alpha$ BAO (see Fig.~\ref{fig:bao-sne}); as SDSS shows some internal tension between $D_{\rm M}/r_{\rm d}$ and $D_{\rm H}/r_{\rm d}$ around these redshifts, the DESI preference currently appears more reliable.

The location of the reconstructed maxima can be traced to the distance residuals shown in Fig.~\ref{fig:bao-sne}. The distances measured by all datasets except DES-Dovekie show deficits relative to \textit{Planck} $\Lambda$CDM around $z\in[0.6,0.8)$. This suggests a preference for increased dark energy density in the same redshift range when considering $D_{\rm H}/r_{\rm d}$ and $D_{\rm V}/r_{\rm d}$ measurements, which constrain the Hubble rate directly, or at slightly lower redshifts when considering $\mu$ and $D_{\rm M}/r_{\rm d}$ (and possibly also $D_{\rm V}/r_{\rm d}$), which involve an integral over $H(z)$. By contrast, DES-Dovekie shows a milder downward fluctuation at slightly higher redshifts, $z\sim0.8$--$1.6$, where there are few SNe~Ia measurements, shifting the reconstructed maximum towards the adjacent $z\in[0.8,1.1)$ bin when combined with SDSS.

From a physical perspective, a local maximum in $f_{\rm DE}(z)$ is particularly interesting. As discussed in Sec.~\ref{sec:methodology}, Eq.~\eqref{eq:fde-deriv} implies that extrema in the dark energy density correspond to crossings of the phantom divide, $w_{\rm DE}=-1$. The reconstructed maxima in Fig.~\ref{fig:fde-posteriors} suggest a transition from $w_{\rm DE}>-1$ at $z\lesssim0.6$ to $w_{\rm DE}<-1$ at $z\gtrsim0.8$. While the statistical significance of this feature depends on the dataset combination, its repeated appearance across all combinations indicates a shared preference robust to potential systematics in any one dataset alone.

Comparing with the CPL parametrization, whose posteriors are represented by the light green bands in Fig.~\ref{fig:fde-posteriors}, highlights the differences between the piecewise-constant reconstructions and smooth low-dimensional dark energy models. In general, the reconstructed $f_{\rm DE}(z)$ peaks at slightly higher redshifts than preferred by CPL, and the bin containing the local maximum often lies above the CPL $1\sigma$ posterior. Outside of this localized feature, these methods yield $f_{\rm DE}(z)$ posteriors consistent within approximately $1\sigma$. This suggests that current data cannot strongly identify local deviations in the dark energy density beyond the average evolutionary trend.

\subsection{\label{sec:wde-res}Equation of State}

\begin{figure*}
    \centering
    \includegraphics[width=\linewidth]{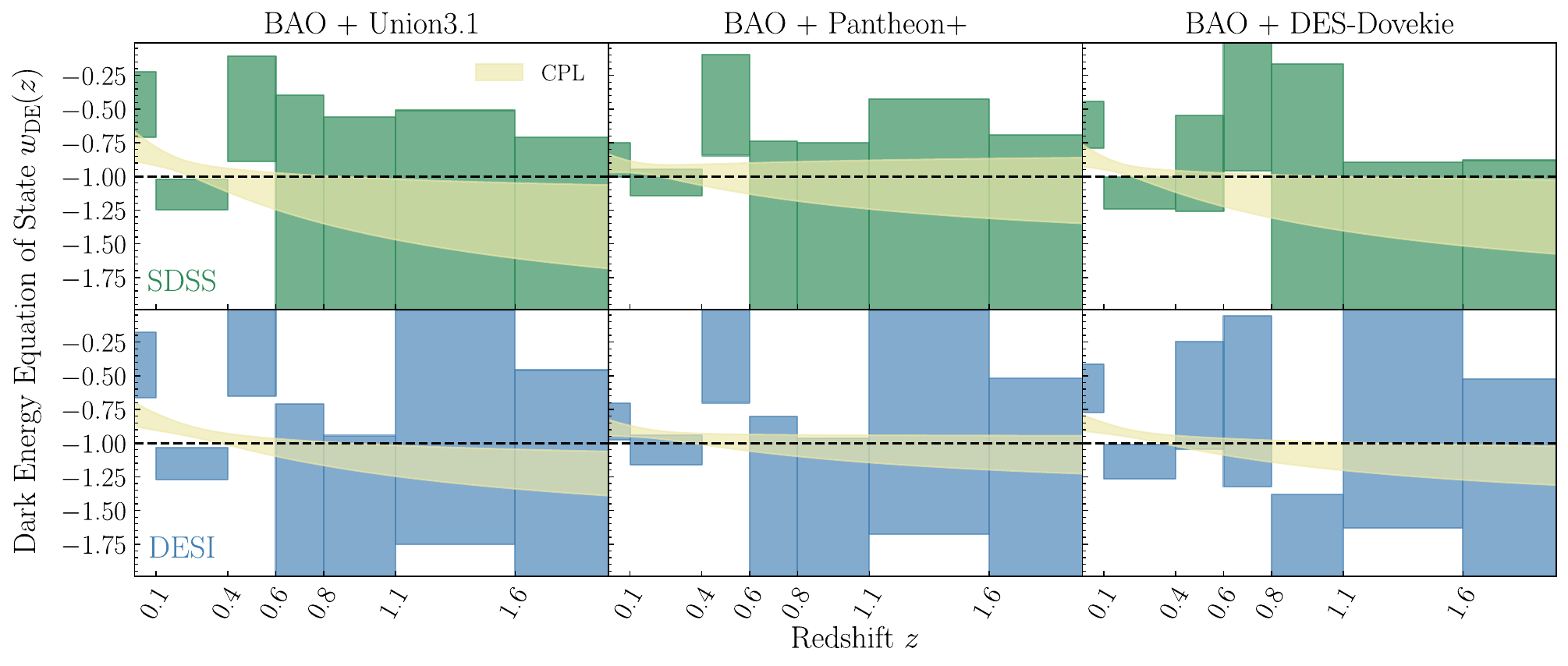}
    \caption[]{Piecewise-constant reconstructions ($1\sigma$ posteriors) of the dark energy equation of state, $w_{\rm DE}(z)$, for each BAO and SNe~Ia combination, calibrated by a BBN prior on $\Omega_bh^2$ and a conservative CMB prior on $\theta_*$ in each case. The top row shows combinations containing BAO from SDSS (green), while the bottom row shows combinations containing DESI BAO (blue). The light green bands represent the $1\sigma$ constraints obtained from the CPL parametrization of $w_{\rm DE}(z)$. The first bin, $0\leq z<0.1$, consistently prefers quintessence-like values, $w_{\rm DE}>-1$. The reconstructions show a marginal preference for an upward fluctuation near $z\in[0.4,0.6)$ followed by a subsequent decrease between $z\in[0.8,1.1)$. Hence, a phantom crossing near $z\sim0.6$--$0.8$ is faintly preferred, consistent with the local maxima observed in the $f_{\rm DE}(z)$ reconstructions (Fig.~\ref{fig:fde-posteriors}).}
    \label{fig:wde-posteriors}
\end{figure*}

Figure~\ref{fig:wde-posteriors} shows the reconstructed dark energy equation of state, $w_{\rm DE}(z)$, for each combination of BAO and SNe~Ia datasets, calibrated by a BBN prior on $\Omega_b h^2$ and a conservative CMB prior on $\theta_*$.

The reconstructions exhibit qualitatively similar oscillatory features across the different datasets, with comparable statistical significance between DESI and SDSS combinations and modest variations between SNe~Ia catalogs. The lowest-redshift bin contains only SNe~Ia measurements and always prefers quintessence-like values, $w_{\rm DE}>-1$, at around $1$--$2\sigma$ significance. While the smallest such deviation occurs with Pantheon+, we note that Ref.~\cite{Hoyt:2026fve} identified a potential low-redshift systematic in this catalog that has not been accounted for in the public Pantheon+ measurements used in this paper. At intermediate redshifts, most combinations show a second fluctuation above $w_{\rm DE}=-1$ at $\sim1$--$2\sigma$, followed by a return towards or slightly below this line at higher redshifts. These last two features faintly indicate a phantom crossing, in agreement with the local maxima identified in the energy density reconstructions (Fig.~\ref{fig:fde-posteriors}).

These oscillatory features are apparently related to the distance residuals shown in Fig.~\ref{fig:bao-sne}. The preference for $w_{\rm DE}>-1$ at low redshifts tracks the initial downward trend in the SNe~Ia distance measurements relative to \textit{Planck} $\Lambda$CDM. The subsequent upward and marginal downward fluctuations occurring around $z\in[0.4,0.6)$ and $z\in[0.8,1.1)$ in the reconstructions, which are responsible for the phantom crossing feature, are traced by opposing fluctuations in $D_{\rm H}/r_{\rm d}$ at similar redshifts and in the integrated distance measurements ($D_{\rm M}/r_{\rm d}$ and $\mu$) at slightly higher redshifts. An alternative explanation for the oscillations in the reconstructions that we considered is the preference for anti-correlations between adjacent $w_{\rm DE}(z)$ bins, shown in Fig.~\ref{fig:wde-triangle}. However, as these anti-correlations also appear in our reconstructions on mock datasets, which contain no oscillations (Fig.~\ref{fig:mock-validations}), we can exclude them as the sole explanation for these features.

Comparing again with the CPL parametrization, whose posterior is restricted to a uniformly increasing trend (Fig.~\ref{fig:wde-posteriors}), emphasizes the increased flexibility of the reconstructions. The reconstructions show a marginal preference for two separate deviations from $\Lambda$CDM: rising above $w_{\rm DE}=-1$ at low redshifts ($z<0.1$) and, with apparently lower statistical significance, crossing the phantom divide around $z\sim0.6$--$0.8$. Similar patterns appear in recent dark energy reconstructions with comparable amounts of freedom~\cite{Ye:2024ywg,Berti:2025phi,DESI:2025wyn,Ormondroyd:2025iaf,Ye:2026yqk}. The CPL parametrization projects these deviations onto a smooth evolutionary history, preferring $w_0>-1$ at the present day and a phantom crossing around the same redshifts as the reconstructions (Fig.~\ref{fig:wde-posteriors}). Given that the reconstructions never deviate from the CPL posterior by more than $\sim1$--$2\sigma$, we cannot interpret these apparent oscillations as evidence for new dark energy phenomenology. However, the recurrence of these features warrants further study of their combined statistical significance, which we will address in an upcoming paper, and motivates revisiting them with future datasets.

\subsection{\label{sec:extensions-robustness}Extensions and Robustness}

\begin{figure*}
    \centering
    \includegraphics[width=\linewidth]{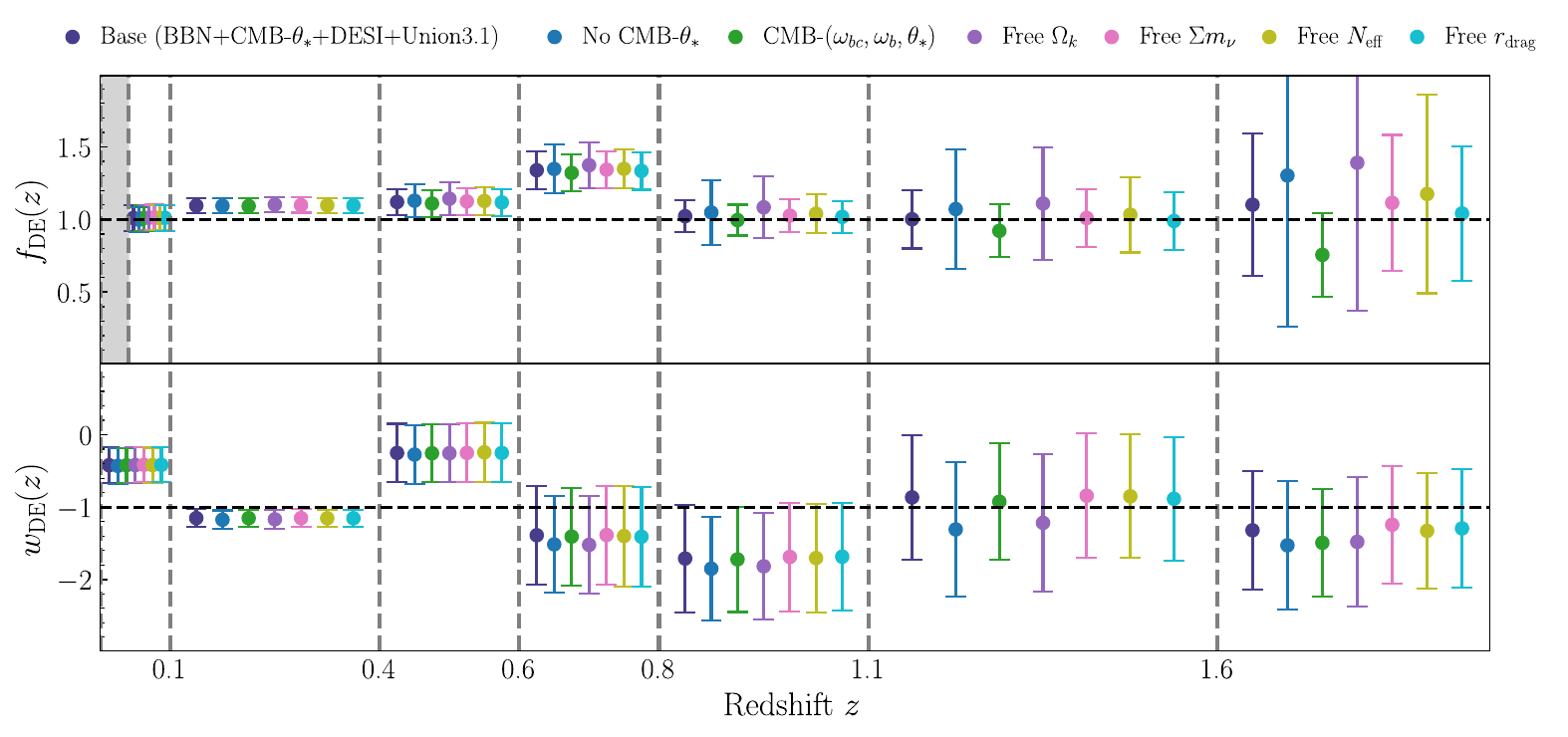}
    \caption{
        Robustness tests of the dark energy density and equation of state reconstructions obtained from our chosen baseline dataset, DESI and Union3.1, calibrated by BBN and CMB-$\theta_*$ priors. The gray dashed lines show the reconstructed bins' redshift ranges, and the interior points correspond to the constraints obtained from reconstructions with one change each relative to the baseline case. The changes are, from left to right: (1) removing the CMB prior on $\theta_*$, (2) replacing the BBN and CMB-$\theta_*$ priors with a correlated Gaussian prior on $(\Omega_{bc}h^2,\Omega_bh^2,\theta_*)$ from early-time CMB anisotropies~\cite{Lemos:2023xhs}, (3--5) allowing the spatial curvature $\Omega_k$, neutrino mass sum $\sum m_\nu$, and effective number of relativistic species $N_{\rm eff}$ to vary freely, one parameter at a time, and (6) treating the sound horizon $r_{\rm d}$ as a free parameter, following the methodology of Ref.~\cite{GarciaEscudero:2025lef}. All variations have only marginal effects on the reconstructions, indicating that the main features are robust to our fiducial parameter assumptions.
    }
    \label{fig:extensions}
\end{figure*}

While the reconstructions in Secs.~\ref{sec:fde-res} and~\ref{sec:wde-res} are qualitatively consistent across the different BAO and SNe~Ia combinations, their common features may arise from assumptions shared by all analyses rather than from an underlying preference in the data. To assess this possibility, we perform a series of robustness tests on the dataset combination that shows the largest deviations from $\Lambda$CDM: DESI BAO combined with Union3.1 SNe~Ia, calibrated by BBN and conservative CMB-$\theta_*$ priors.

The following variations are considered, one at a time:
\begin{enumerate}
    \item Removing the CMB-$\theta_*$ prior, testing whether the reconstruction is driven by acoustic-scale information inherited from the CMB, which depends on assumptions about the early Universe.\footnote{For an example of alternative assumptions that significantly modify the CMB inference of $\theta_*$, see self-interacting neutrinos~\cite{Brinckmann:2020bcn}.}
    \item Incorporating additional CMB information through a correlated Gaussian prior on $\Omega_bh^2$, $\theta_*$, and the total baryonic and cold dark matter density, $\Omega_{bc}h^2$, from early-time anisotropies~\cite {Lemos:2023xhs}, replacing our baseline BBN and CMB-$\theta_*$ priors. In principle, this reintroduces the $\Omega_m$~tension between CMB and late-time distance measurements.
    \item Allowing the spatial curvature parameter $\Omega_k$ to vary with a uniform prior $\mathcal U(-0.3,0.3)$, testing the effect of assuming spatial flatness.
    \item Allowing the neutrino mass sum, $\sum m_\nu$, to vary with a uniform prior $\mathcal U(0,5)\,{\rm eV}$, freeing this parameter from the fiducial value $\sum m_\nu=0.06\,{\rm eV}$, which lacks strong physical motivation.
    \item Allowing the effective number of relativistic species, $N_{\rm eff}$, to vary from the standard model value $N_{\rm eff}=3.044$. Following Ref.~\cite{DESI:2025zgx}, we impose a correlated Gaussian prior on $N_{\rm eff}$ and $\Omega_bh^2$ derived from \texttt{PRyMordial}~\cite{Schoneberg:2024ifp,Burns:2023sgx}.
    \item Promoting the sound horizon at baryon drag, $r_d$, to a free parameter, following the methodology of Ref.~\cite{GarciaEscudero:2025lef}, rather than using the approximation in Eq.~\eqref{eq:rdrag}. This reduces dependence on the assumed $\Lambda$CDM recombination history, although some of this dependence persists in the CMB-$\theta_*$ prior.
\end{enumerate}

The results are shown in Fig.~\ref{fig:extensions}. In all cases, the reconstructed $f_{\rm DE}(z)$ and $w_{\rm DE}(z)$ remain consistent with the baseline reconstruction within their statistical uncertainties.

The main features of the $f_{\rm DE}(z)$ reconstructions, namely deviations from $\Lambda$CDM occurring below $z\lesssim0.8$, are robust to the modifications discussed above. The CMB prior on $(\Omega_{bc}h^2,\Omega_bh^2,\theta_*)$ moderately decreases the reconstruction uncertainties at higher redshifts, $z\gtrsim0.4$, as expected from the additional information on the matter density and the less conservative sound-horizon-scale prior. Conversely, removing the $\theta_*$ prior or allowing spatial curvature to vary slightly broadens the reconstructed posteriors around the same redshifts, reflecting the increased geometric freedom in the background expansion history. Saliently, these changes primarily affect the size of the uncertainties rather than the reconstructed shape itself; in particular, the local maximum around $z\sim0.6$--$0.8$ remains present in all cases.

The reconstructed equation of state is significantly more stable under these variations. Both the location and amplitude of the oscillatory features observed in Fig.~\ref{fig:wde-posteriors} remain nearly unchanged across all extensions considered. This overall stability indicates that these reconstructions are primarily driven by the BAO and SNe~Ia distance measurements rather than by any specific parameter assumptions adopted in the baseline analysis.

\subsection{\label{sec:model-comparison}Model Comparison}

\begin{figure*}[t]
    \centering
    \includegraphics[width=\linewidth]{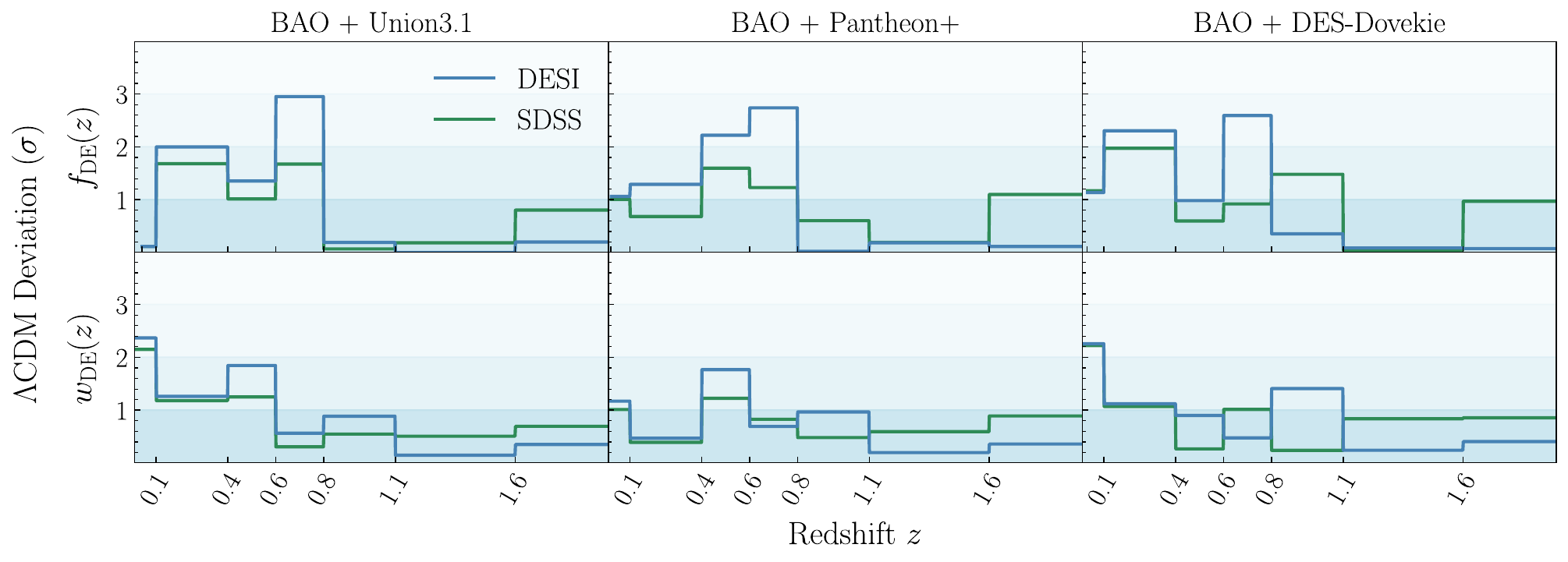}
    \caption{Local deviations from $\Lambda$CDM expectations, $f_{\rm DE}=1$ and $w_{\rm DE}=-1$, expressed as equivalent Gaussian $\sigma$ significances (Sec.~\ref{sec:statistics}). Results from the $f_{\rm DE}(z)$ and $w_{\rm DE}(z)$ reconstructions are shown in the top and bottom rows, respectively. The columns correspond to the SNe~Ia dataset included alongside BAO from SDSS (green lines) and DESI (blue lines). When DESI is included, $f_{\rm DE}(z)$ deviates by $\sim2.6$--$3\sigma$ between $z\in[0.6,0.8)$, where these reconstructions exhibit a local maximum (Fig.~\ref{fig:fde-posteriors}). This significance is reduced and sometimes shifted to adjacent bins when DESI is replaced by SDSS. By contrast, $w_{\rm DE}(z)$ deviates most strongly ($\sim1.1$--$2.4\sigma$) between $z\in[0,0.1)$, preferring $w_{\rm DE}>-1$ (Fig.~\ref{fig:wde-posteriors}), regardless of the BAO dataset included.}
    \label{fig:local-deviations}
\end{figure*}

\begin{figure*}
    \centering
    \includegraphics[width=\linewidth]{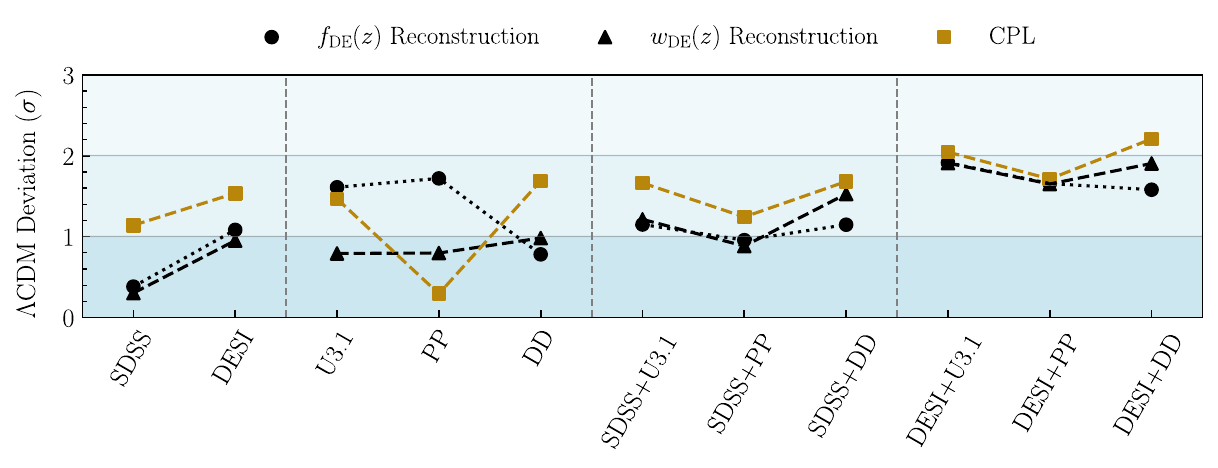}
    \caption{Global deviations from $\Lambda$CDM expressed as equivalent Gaussian $\sigma$ significances, calculated from the total $\Delta\chi^2$ and the difference in degrees of freedom (Sec.~\ref{sec:statistics}). Deviations are shown for the $f_{\rm DE}(z)$ and $w_{\rm DE}(z)$ reconstructions alongside the CPL parametrization of $w_{\rm DE}(z)$ for each BAO and SNe~Ia dataset (see the corresponding reconstructions in Fig.~\ref{fig:individual-datasets}) and their combinations (Figs.~\ref{fig:fde-posteriors} and \ref{fig:wde-posteriors}). The individual datasets and the combinations including SDSS are consistent with $\Lambda$CDM at below $2\sigma$. When DESI is included, deviations around $1.5$--$2\sigma$ occur regardless of the SNe~Ia dataset or dark energy methodology. The fact that the reconstructions provide significances similar to CPL suggests that they offer sufficient improvements in fitting these datasets to justify their additional freedom.}
    \label{fig:global-deviations}
\end{figure*}

\begin{table*}[t]
    \centering
    \setlength{\tabcolsep}{12pt}
    \begin{tabular}{lcccccc}
    \toprule
    & \multicolumn{3}{c}{{DESI+}} & \multicolumn{3}{c}{{SDSS+}} \\
    \cmidrule(lr){2-4} \cmidrule(lr){5-7} 
     & {Union3.1} & {Pantheon+} & {DES-Dovekie} & {Union3.1} & {Pantheon+} & {DES-Dovekie} \\ 
    \midrule
    $w_{\rm DE}(z)\quad\sigma_{\rm tension}$ & $1.2 \pm 0.2$ & $1.2 \pm 0.2$ & $2.0 \pm 0.3$ & $1.1 \pm 0.2$ & $1.1 \pm 0.1$ & $1.2 \pm 0.1$ \\
    $w_{\rm DE}(z)\quad\log B$ & $-3.6 \pm 0.8$ & $-5.2 \pm 0.8$ & $-4.2 \pm 0.7$ & $-3.7 \pm 0.6$ & $-5.7 \pm 0.6$ & $-4.8 \pm 0.6$ \\
    \bottomrule
    \end{tabular}
    \caption{Suspiciousness tension metric, $\sigma_{\rm tension}$, and Bayes factor with respect to $\Lambda$CDM, $\log B$, for the $w_{\rm DE}(z)$ reconstructions and each BAO-SNe~Ia combination (Sec.~\ref{sec:statistics}). All dataset combinations are internally consistent at the $1\sigma$ level, apart from DESI+DES-Dovekie, which shows a mild $2\sigma$ internal tension. The reconstructions perform worse than $\Lambda$CDM according to the Bayesian evidence, as expected from their intentionally broad prior volumes; see the discussion in the main text.}
    \label{tab:bayesian-stats}
\end{table*}

In this section, we quantify the statistical significance of the departures from $\Lambda$CDM identified in the previous two sections and assess the internal consistency of the reconstructions.

Local deviations from $\Lambda$CDM expectations, $f_{\rm DE}=1$ and $w_{\rm DE}=-1$, are evaluated for each reconstruction bin and expressed as equivalent Gaussian significances in Fig.~\ref{fig:local-deviations}. Overall, the $f_{\rm DE}(z)$ reconstructions display larger deviations than the corresponding $w_{\rm DE}(z)$ results, as the dark energy density is more directly constrained by the data (Sec.~\ref{sec:background}). In almost all cases, the apparent local maximum identified in the $f_{\rm DE}(z)$ reconstructions represents the most significant deviation ($\sim2.6$--$3\sigma$ when DESI is included), followed by the indications for $f_{\rm DE}>1$ at lower redshifts. Replacing DESI with SDSS reduces the significances by $\sim0.4$--$1.4\sigma$, sometimes shifting the preference to adjacent redshift intervals. The largest deviations in $w_{\rm DE}(z)$ occur for $z<0.1$, where quintessence-like values are preferred at $\sim1$--$2.4\sigma$ significance. The equation-of-state bins that drive the preference for a phantom crossing, $z\in[0.4,0.6)$ and $z\in[0.8,1.1)$, often show lower significances and greater dependence on the datasets used.

The total deviation from $\Lambda$CDM is measured by the minimum chi-square difference, $\Delta\chi^2$, between the reconstructions and $\Lambda$CDM fits to the same datasets. Because $\Lambda$CDM is nested within our reconstructions, this difference approximately follows a chi-squared distribution with degrees of freedom equal to the parameter difference $p_{\rm recon}-p_{\Lambda\mathrm{CDM}}=7$, permitting a translation to Gaussian significances. These global deviations are shown in Fig.~\ref{fig:global-deviations} alongside the corresponding significances obtained from the CPL parametrization with $p_{\rm CPL}-p_{\Lambda\mathrm{CDM}}=2$. The significances are overall more modest than those identified locally (Fig.~\ref{fig:local-deviations}), as the reconstructions are penalized by additional degrees of freedom, not all of which contribute to improving the fit. Individual BAO or SNe~Ia datasets, alongside SNe~Ia combinations with SDSS, are consistent with $\Lambda$CDM at below $2\sigma$ significance, while combinations including DESI more consistently yield deviations of order $\sim1.6$--$2\sigma$, largely independent of the SNe~Ia sample or dark energy methodology. This point is worth emphasizing: the reconstruction significances are comparable to those of CPL despite being penalized by five additional parameters, suggesting that the reconstructions compensate for their added freedom with a sufficient improvement in fitting the BAO and SNe~Ia distances.

Bayesian statistics provide a complementary metric for the performance of the reconstructions relative to $\Lambda$CDM and additional information on their internal consistency, incorporating the full posterior distribution. Bayes factors relative to $\Lambda$CDM and suspiciousness statistics are reported for the $w_{\rm DE}(z)$ reconstructions in Table~\ref{tab:bayesian-stats}.\footnote{We report these statistics for $w_{\rm DE}(z)$ rather than $f_{\rm DE}(z)$ for two reasons. First, our $f_{\rm DE}(z)$ nested sampling runs struggled to find an initial point when sampling over $\theta_*$, and sampling over $h$ instead has large effects on the resulting Bayesian evidences. Second, calculating the suspiciousness requires separate runs on the BAO and SNe~Ia datasets, which each required separate values of $z_{\min}$ when reconstructing the energy density. The $w_{\rm DE}(z)$ reconstructions avoid both complications.} The suspiciousness points towards mutual consistency for every BAO and SNe~Ia combination apart from DESI and DES-Dovekie, which shows a mild $\sim2\sigma$ internal tension, perhaps due to these datasets' opposing distance preferences around $z\in[0.6,0.8)$ (see Fig.~\ref{fig:bao-sne}). The absence of strong internal inconsistencies between most dataset combinations, which all show common preferences in Figs.~\ref{fig:fde-posteriors} and \ref{fig:wde-posteriors}, suggests that these preferences result from coherent trends shared between the late-time measurements rather than from a single discrepant probe.

The Bayes factors generally disfavor the reconstructions relative to $\Lambda$CDM. However, as discussed in Sec.~\ref{sec:statistics} and Footnote~\ref{footnote:bayes-factor}, this is expected from the broad prior volumes of the reconstructions, which are intended to avoid strong prejudices and fully explore the phenomenology allowed by current data, rather than to define a minimal predictive framework.

\subsection{\label{sec:lpca}Localized Principal Component Analysis}

\begin{figure*}
    \centering
    \includegraphics[width=\linewidth]{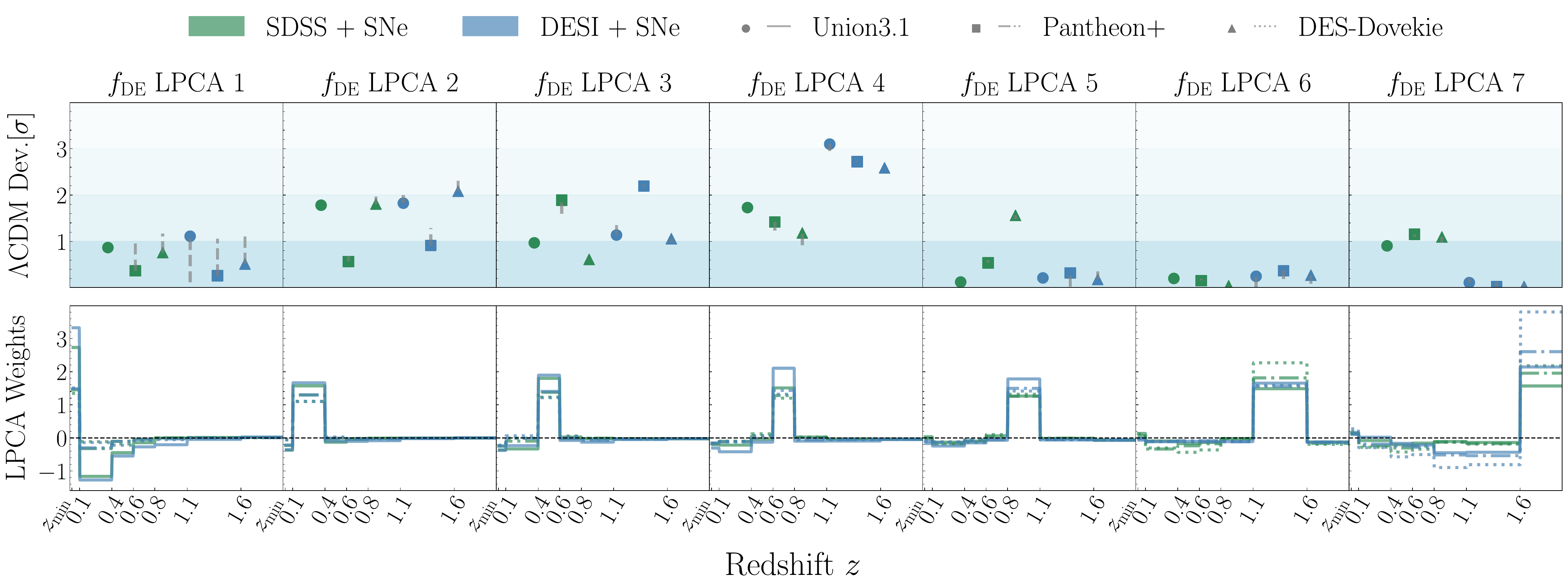}
    \caption{The results of localized principal component analysis (LPCA) on the $f_{\rm DE}(z)$ reconstructions for each BAO-SNe~Ia combination (Fig.~\ref{fig:fde-posteriors}). The bottom row shows the LPCA weights: the contributions from each redshift range to the principal component amplitudes. The points in the upper row show the uncorrelated deviations of these amplitudes from $\Lambda$CDM in Gaussian $\sigma$ significance, while the attached dashed lines show the shifts of these points from the corresponding correlated deviations (Fig.~\ref{fig:local-deviations}). Colors denote the BAO dataset included, green for SDSS and blue for DESI, while the markers and line styles denote the SNe~Ia sample used in combination (see legend). A robust deviation is seen in the fourth principal component, whose weights are centered around the local maximum of the $f_{\rm DE}(z)$ reconstructions (Fig.~\ref{fig:fde-posteriors}).}
    \label{fig:fde-lpca}
\end{figure*}

\begin{figure*}
    \centering
    \includegraphics[width=\linewidth]{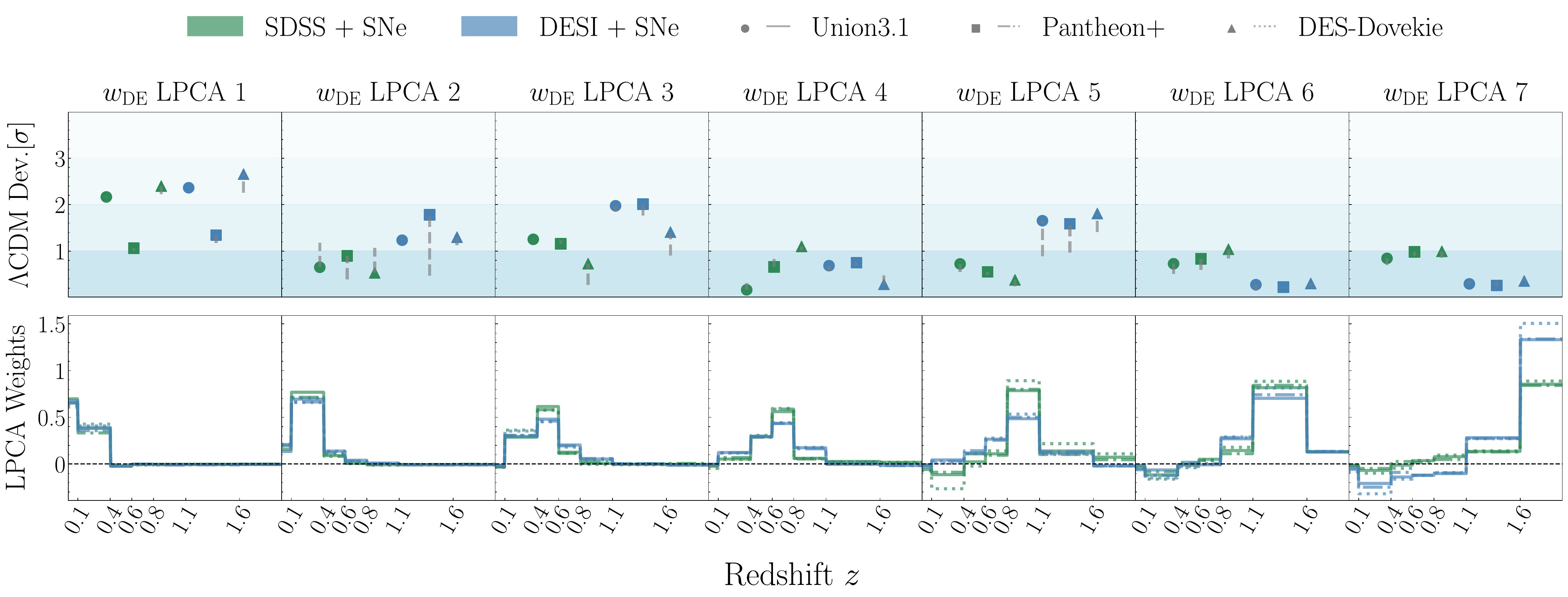}
    \caption{The results of localized principal component analysis (LPCA) on the $w_{\rm DE}(z)$ reconstructions for each BAO-SNe~Ia combination (Fig.~\ref{fig:wde-posteriors}). The bottom row shows the LPCA weights: the contributions from each redshift range to the principal component amplitudes. The points in the upper row show the uncorrelated deviations of these amplitudes from $\Lambda$CDM in Gaussian $\sigma$ significance, while the attached dashed lines show the shifts of these points from the corresponding correlated deviations (Fig.~\ref{fig:local-deviations}). Colors denote the BAO dataset included, green for SDSS and blue for DESI, while the markers and line styles denote the SNe~Ia sample used in combination (see legend). The first principal component shows relatively large but inconsistent deviations, corresponding to the low-redshift preference for $w_{\rm DE}>-1$ in the direct reconstructions. In DESI combinations, three more mild deviations appear, including two around the possible phantom crossing in Fig.~\ref{fig:wde-posteriors}.}
    \label{fig:wde-lpca}
\end{figure*}

The results of applying localized principal component analysis to the reconstructions are summarized in Figs.~\ref{fig:fde-lpca} and \ref{fig:wde-lpca}. In these figures, the bottom rows show the redshift weights of the principal components, while the upper rows show the significances of these components' deviations from $\Lambda$CDM. Also in the upper rows, the vertical dashed lines represent the difference between the corresponding correlated local deviations found in Fig.~\ref{fig:local-deviations} and the \textit{un}correlated deviations identified by LPCA. The fact that these differences are almost always small, and sometimes show an increase in significance under LPCA, suggests that the individual deviations from $\Lambda$CDM found in our reconstructions of the dark energy density and equation of state (Figs.~\ref{fig:fde-posteriors} and \ref{fig:wde-posteriors}, respectively) are more likely to represent genuinely independent features of the data.

Fig.~\ref{fig:fde-lpca} contains the $f_{\rm DE}(z)$ results. While the statistical deviations from $\Lambda$CDM of the first three principal components depend considerably on the SNe~Ia dataset used, the fourth principal component is much more consistent in this regard. The fourth component displays the largest deviations from $\Lambda$CDM: around $2.6$--$3.2\sigma$ in DESI combinations, significantly greater than the $\sim1$--$2\sigma$ obtained with SDSS. The weights for this amplitude are relatively well confined to the range $z\in[0.6,0.8)$, approximately matching the location of the local maximum in the $f_{\rm DE}(z)$ reconstructions (Fig.~\ref{fig:fde-posteriors}).

Fig.~\ref{fig:wde-lpca} shows analogous results for the equation of state. Here, the weights are generally broader in redshift, reflecting the fact that $w_{\rm DE}(z)$ is less directly constrained by distance observables (Sec.~\ref{sec:background}). Consistent with the energy density results, which in principle depend on the redshift integral of the equation of state, the most notable deviations from $\Lambda$CDM occur in the first three principal components of $w_{\rm DE}(z)$, whose weights span roughly $z\in[0,1.1)$. The first component shows the largest deviations, $\sim1$--$2.6\sigma$, but exhibits appreciable scatter across SNe~Ia datasets. When DESI is included, three more $1$--$2\sigma$ deviations are seen. These notably include the third and fifth principal components, whose weights surround the location of the phantom crossing weakly preferred by the direct reconstructions (Fig.~\ref{fig:wde-posteriors}).

\section{\label{sec:conclusions}Conclusions}

In this work, we performed minimal reconstructions of dark energy at redshifts $z<4.2$ using the latest publicly available baryon acoustic oscillation and Type~Ia supernovae datasets. By constraining piecewise-constant parametrizations of both the normalized dark energy density, $f_{\rm DE}(z)$, and the dark energy equation of state, $w_{\rm DE}(z)$, we inferred the approximate average values of these functions over seven redshift bins, matching the redshift ranges of the galaxy populations used in the DESI BAO analysis plus an additional bin for $z<0.1$ to capture the preferences of low-redshift SNe~Ia. Our baseline reconstructions are calibrated using only a Big Bang Nucleosynthesis prior on the physical baryon density and a conservative prior on the angular sound horizon at recombination from CMB data, the least model-dependent geometric quantity constrained by such observations. By including only this background-level information, we avoid other potential causes for the evolving dark energy signal, such as the $\Omega_m$~tension between CMB and late-time probes and systematic uncertainties surrounding the optical depth to reionization.

Our reconstruction methodology was carefully chosen to be data-driven and independent of theoretical prejudices. We avoid enforcing any correlations between our reconstructed parameters, physically motivated or otherwise, leaving only the correlations imposed by the observables themselves. The only freedom in piecewise-constant reconstructions, the number and locations of the redshift bins, is largely removed by aligning our bins with the DESI tracers, ensuring that the reconstructions are fully robust to theoretical uncertainties in the effective redshifts of these measurements.

Our main results are the consistent shapes that emerge for $f_{\rm DE}(z)$ and $w_{\rm DE}(z)$ when these functions are reconstructed from combinations of widely used BAO and SNe~Ia datasets. This consistency suggests that the evolving dark energy signal is caused by genuine features in late-time distance measurements, together with our core cosmological assumptions, rather than by individual systematics or random fluctuations. The evolutionary histories found in our baseline analysis are robust to numerous changes and extensions: removing the CMB acoustic-scale prior or adding additional CMB information, freely varying parameters previously held fixed to fiducial values (the spatial curvature, neutrino mass sum, and effective number of relativistic species), and relaxing assumptions about the early-Universe recombination history by treating the sound horizon $r_d$ as a free parameter. The robustness of these reconstructed histories is further supported by the results of localized principal component analysis, which yielded uncorrelated amplitudes for $f_{\rm DE}(z)$ and $w_{\rm DE}(z)$ whose deviations from $\Lambda$CDM reproduce the patterns found in the direct reconstructions. This suggests that the shapes of the reconstructions are not driven by a subset of discrepant data points whose effects are propagated across redshifts by the intrinsic correlations of the reconstructed parameters.

The largest deviations from $\Lambda$CDM are found in the normalized energy density around $z\sim0.6$--$0.8$, where this function appears to reach a local maximum. This deviation has an equivalent Gaussian significance of $\sim2.6$--$3\sigma$ in BAO-SNe~Ia combinations including DESI and $\sim1.5\sigma$ in those including SDSS, where the preference sometimes shifts to adjacent redshift intervals. If genuine, such a local extremum would provide direct evidence for a phantom crossing in the dark energy equation of state around these redshifts. We defer exploring the statistical significance of this feature to future work, although see Refs.~\cite{Ozulker:2025ehg, Keeley:2025rlg} for current research in this direction. At lower redshifts, the $f_{\rm DE}(z)$ reconstructions generally decline towards the $\Lambda$CDM limit, showing additional preferences for $f_{\rm DE}>1$ at around the $2\sigma$ level with DESI and $1$--$2\sigma$ with SDSS. Fewer reliable deviations are found above $z\gtrsim0.8$, where values near the cosmological constant limit are preferred.

The $w_{\rm DE}(z)$ reconstructions prefer somewhat less statistically significant deviations from $\Lambda$CDM, as the dark energy equation of state is less directly constrained by cosmological distance measurements. Nevertheless, these reconstructions show a common pattern of deviations from $\Lambda$CDM across all BAO and SNe~Ia combinations considered: (\textit{i}) a preference for $w_{\rm DE}>-1$ in the lowest-redshift bin, and (\textit{ii}) a second milder fluctuation above $w_{\rm DE}=-1$ before $z\lesssim0.6$, followed by a fluctuation below this line after $z\gtrsim0.8$, tracing the preference for a phantom crossing between $z\sim0.6$--$0.8$ identified by the energy density reconstructions. The first deviation is primarily driven by the SNe~Ia observations and is found with $\sim1$--$2.4\sigma$ significance in all dataset combinations, while the deviations corresponding to the phantom crossing indication are mostly caused by the BAO measurements (Fig.~\ref{fig:individual-datasets}) and have significances around $0.8$--$1.8\sigma$. Falling mostly below the $2\sigma$ level, these deviations cannot be interpreted as evidence for non-trivial oscillatory dynamics, although their persistence across dataset combinations warrants further study.

Our reconstructions are mostly statistically consistent with the broad trends obtained using the CPL parametrization for $w_{\rm DE}(z)$. However, we find tentative hints for sharper features than this parametrization can capture. In particular, the $f_{\rm DE}(z)$ reconstructions suggest a local maximum that is steeper and located at higher redshifts than suggested by the corresponding CPL posteriors. By averaging over such localized features, low-dimensional parametrizations may in some cases provide less compelling evidence for dark energy evolution than flexible reconstruction methodologies, such as the one presented here.

As we constrained seven dark energy parameters using six separate dataset combinations for both the dark energy density and equation of state, the deviations from $\Lambda$CDM that we find may be affected by the look-elsewhere effect, whereby significant findings occur in a large number of independent trials due to random chance. We expect this effect to be small, as the naive number of $\sim7\times6=42$ trials for the reconstructed functions are in practice highly correlated. However, it is still useful to complement these local statistics with other metrics that incorporate more of the full posterior distribution. From the $\Delta\chi^2$ between the reconstructions and $\Lambda$CDM fits to DESI combinations with SNe~Ia, we find $\sim2\sigma$ support for the seven additional degrees of freedom in the reconstructions. A similar significance is derived from the $\Delta\chi^2$ between CPL and $\Lambda$CDM, suggesting that evidence for evolving dark energy may saturate around this level with current background-level data. The Bayesian evidence disfavors the reconstructions, although this is expected from the fact that our reconstructions do not represent minimal predictive models and have intentionally broad prior volumes. Finally, the Bayesian suspiciousness statistic suggests that all BAO-SNe~Ia combinations are internally consistent, with the slight exception of DESI and DES-Dovekie, which show a mild $\sim2\sigma$ internal tension, possibly owing to the different preferences of these datasets around $z\in[0.6,0.8)$.

Future data from DESI, Euclid, Rubin LSST, Roman, and next-generation supernova surveys will substantially improve the precision and redshift coverage of late-time distance measurements. These observations will determine whether the dark energy behaviors identified here persist with increasing significance or instead converge towards $\Lambda$CDM expectations. In this context, minimal reconstruction methodologies will continue to provide essential robustness tests for standard dark energy models and may identify local phenomenology hidden by these parametrizations.

\acknowledgments
We are grateful to William Luke Matthewson for clarifying details about the piecewise-constant reconstructions in Ref.~\cite{DESI:2025fii}, and to Taylor J. Hoyt and David Rubin for providing the recalibrated Union3.1 data from Ref.~\cite{Hoyt:2026fve} and additional details necessary for producing Fig.~\ref{fig:shared-sne}.
E.D.V. is supported by a Royal Society Dorothy Hodgkin Research Fellowship. L.A.E. acknowledges support from T\"{U}B\.{I}TAK through a postdoctoral researcher fellowship associated with Grant No.~124N627. DH is supported by the Department of Energy under contract DE-SC009193. This article is based upon work from the COST Action CA21136 - ``Addressing observational tensions in cosmology with systematics and fundamental physics (CosmoVerse)'', supported by COST - ``European Cooperation in Science and Technology''. We acknowledge IT Services at The University of Sheffield for the provision of services for High Performance Computing.

\appendix

\section{\label{sec:mocks}Mock Datasets}

\begin{figure*}
    \centering
    \includegraphics[width=\linewidth]{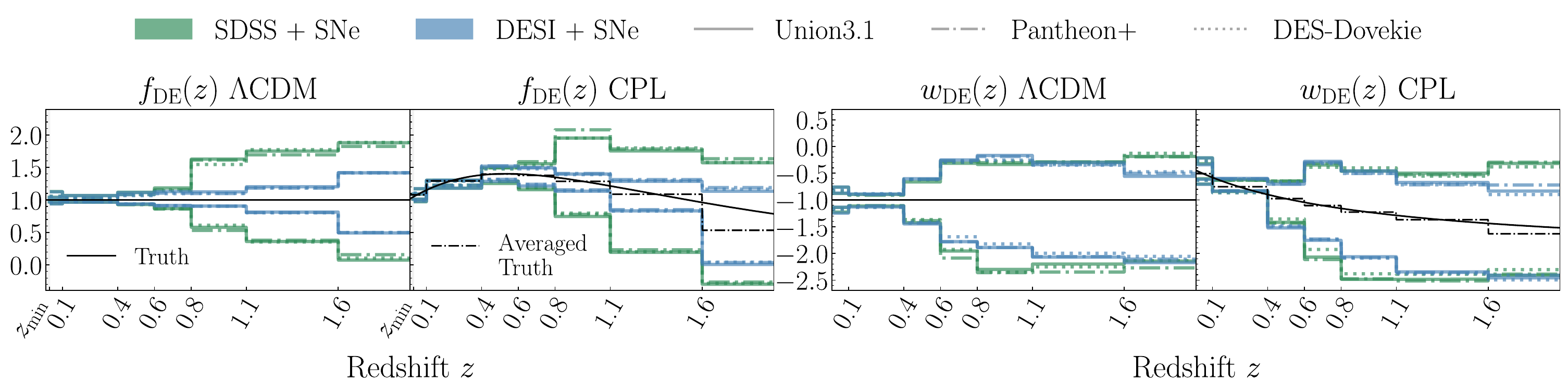}
    \caption{Reconstructions on mock BAO and SNe~Ia datasets generated from fiducial $\Lambda$CDM and CPL cosmologies. Only the boundaries of the reconstructed $1\sigma$ bands are shown for visual clarity. The colors and line styles of the boundaries denote the BAO and SNe~Ia datasets used, respectively. Solid black lines represent the underlying truth, while dash-dot black lines show the truth averaged over the redshift bins of the reconstruction (Eq.~\eqref{eq:g-avg}). In every case, the truth and its binned average are recovered within $1\sigma$.}
    \label{fig:mock-validations}
\end{figure*}

Running our reconstruction methodology on \textit{mock data}, whose mean values come from the predictions of a known cosmological model and whose covariance matrix is inherited from the actual observations, allows us to assess the accuracy of the reconstructions and determine which properties of $f_{\rm DE}(z)$ and $w_{\rm DE}(z)$ they actually recover from the data. For each BAO and SNe~Ia dataset used in this paper, we generated a corresponding mock dataset whose mean values at the measured redshifts come from either a fiducial $\Lambda$CDM model or a fiducial CPL model. We then performed reconstructions on each combination of mock BAO and SNe~Ia datasets (with both datasets generated either from the $\Lambda$CDM cosmology or from the CPL cosmology), always using the same BBN and conservative CMB-$\theta_*$ priors adopted in our reconstructions on real data.

The results are shown in Fig.~\ref{fig:mock-validations}. The true underlying $f_{\rm DE}(z)$ and $w_{\rm DE}(z)$ are recovered well within $1\sigma$ in both the $\Lambda$CDM and CPL cases. In each case, the reconstruction approximately recovers the average of these functions over each redshift bin,
\begin{equation}
    \bar g_i\equiv\frac{1}{z_{i+1}-z_i}\int_{z_i}^{z_{i+1}}\mathrm{d}z\:g(z)\,e_i(z)\,,
    \label{eq:g-avg}
\end{equation}
where $g(z)\in\left\{f_{\rm DE}(z),\,w_{\rm DE}(z)\right\}$, $\left\{z_i\right\}$ are the redshift boundaries of the bins, and $e_i(z)$ is the \textit{i}th top-hat function used in the reconstruction (Eq.~\eqref{eq:g-sum}). Since the mock datasets carry the same uncertainties as the real data, Fig.~\ref{fig:mock-validations} provides some indication of the relative constraining power of these datasets. All SNe~Ia datasets lead to roughly equivalent uncertainties on $f_{\rm DE}(z)$ and $w_{\rm DE}(z)$, whereas DESI BAO leads to much smaller uncertainties than SDSS on $f_{\rm DE}(z)$ at redshifts $z\lesssim0.8$. The choice of BAO dataset has a smaller effect on the $w_{\rm DE}(z)$ reconstructions, as this function is less directly constrained than $f_{\rm DE}(z)$ by BAO measurements.

\begin{figure}
    \centering
    \includegraphics[width=\linewidth]{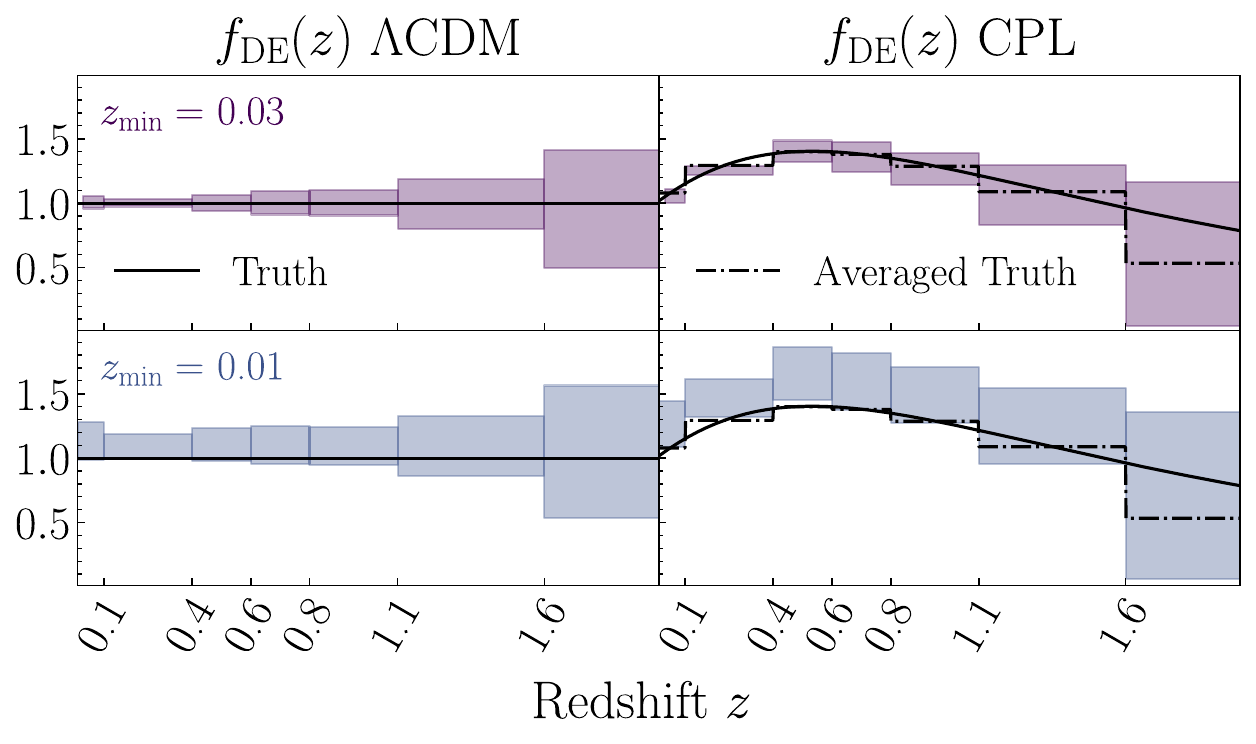}
    \caption{An example of how the minimum redshift, $z_{\rm min}$, of the $f_{\rm DE}(z)$ reconstructions affects the marginalized posteriors, demonstrated here by running reconstructions on mock DESI+DES-Dovekie data generated from fiducial $\Lambda$CDM and CPL cosmologies. The upward drift as $z_{\rm min}$ approaches zero reflects the worsening degeneracy between the lowest-redshift $f_{\rm DE}(z)$ bin and the Hubble constant, leading to an unconstrained direction in the likelihood.}
    \label{fig:zmin}
\end{figure}

As mentioned in Sec.~\ref{sec:de-reconstruction}, we chose the minimum redshift, $z_{\rm min}$, of our $f_{\rm DE}(z)$ reconstructions such that the marginalized constraints on $f_{\rm DE}(z)$ recover the known truth when the reconstructions are performed on mock $\Lambda$CDM and CPL data. Fig.~\ref{fig:zmin} shows how decreasing the value of $z_{\rm min}$ can bias the marginalized constraints, leading to an upward drift away from the known truth. As $z_{\rm min}$ approaches zero, the lowest-redshift $f_{\rm DE}(z)$ bin becomes increasingly degenerate with the Hubble parameter, and the $f_{\rm DE}(z)$ constraints can move along this unconstrained direction of the likelihood without penalty. The value of $z_{\rm min}$ at which this degeneracy leads to significant bias depends on the SNe~Ia dataset included in the analysis, as these datasets provide different amounts of constraining power at $z\lesssim0.1$. For each SNe~Ia dataset, we performed tests analogous to those shown in Fig.~\ref{fig:zmin} and chose $z_{\rm min}$ such that the known truths are approximately centered within the $1\sigma$ bands of the reconstruction posteriors. We also verified that these $z_{\rm min}$ choices are not fine-tuned to the parameter values of our fiducial cosmologies: the same choices lead to unbiased constraints on both steeper and shallower CPL $f_{\rm DE}(z)$ functions. Furthermore, similarly unbiased constraints are obtained when SDSS replaces DESI as the mock BAO dataset.

\section{Individual Datasets}

\begin{figure*}
    \centering
    \includegraphics[width=\linewidth]{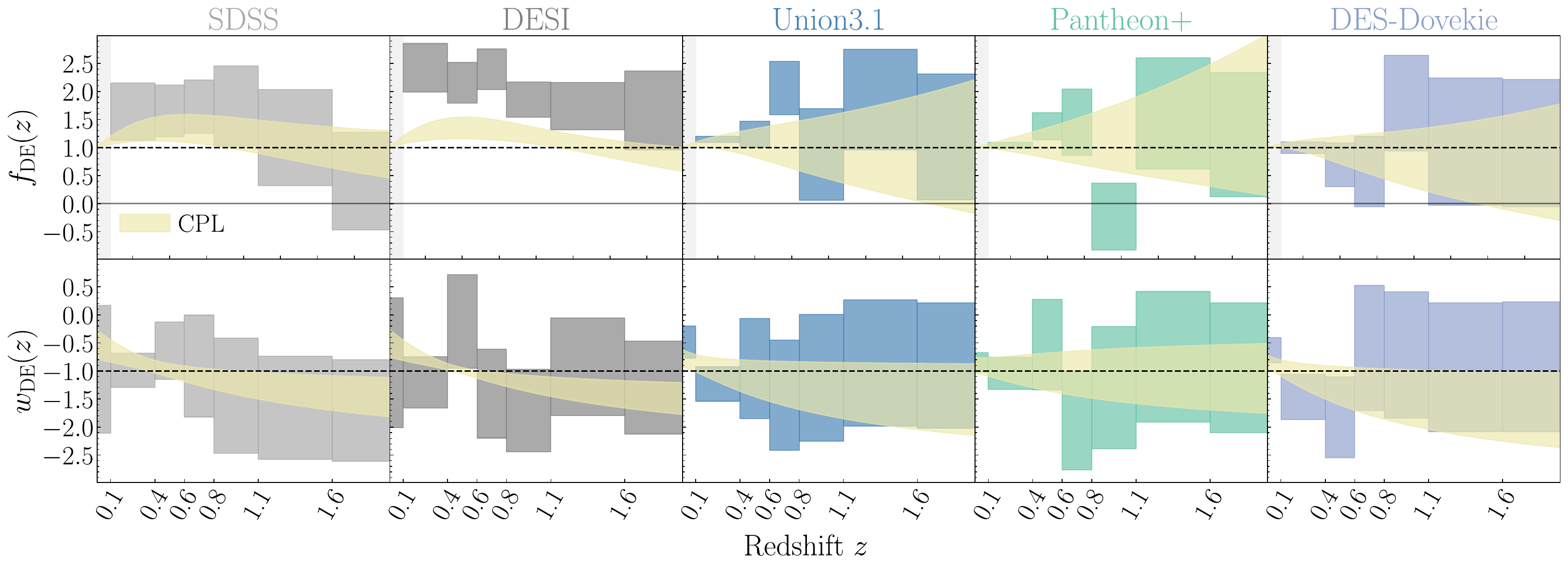}
    \caption{Reconstructions on each individual BAO and SNe~Ia dataset. The BAO-only marginalized $f_{\rm DE}(z)$ posteriors show an upward bias, resulting from the BAO datasets' sparse low-redshift measurements and their inability to break the $f_{\rm DE}$--$h$ degeneracy. The local maximum in $f_{\rm DE}(z)$ seen in dataset combinations (Fig.~\ref{fig:fde-posteriors}) is partially present in the SNe~Ia reconstructions. The pattern of deviations in the main $w_{\rm DE}(z)$ reconstructions (Fig.~\ref{fig:wde-posteriors}) may arise from the tentative preference of the BAO datasets for a phantom crossing between $0.6\leq z<0.8$ together with the SNe~Ia indication for $w_{\rm DE}>-1$ between $0\leq z<0.1$.}
    \label{fig:individual-datasets}
\end{figure*}

In this Appendix, we perform reconstructions using each BAO and SNe~Ia dataset individually to understand their shared preferences and differences. The results are presented in Fig.~\ref{fig:individual-datasets}. To allow for direct comparison between the datasets, $f_{\rm DE}(z)$ is reconstructed between the redshift divisions $\left\{0.1,0.4,0.6,0.8,1.1,1.6,4.2\right\}$ while being fixed to unity between $0\leq z<0.1$ in every case. Conversely, $w_{\rm DE}(z)$ is reconstructed between the same redshift divisions but is allowed to vary between $0\leq z<0.1$. Note that reconstructions from individual datasets are more sensitive to potential systematics and noise, so one cannot interpret their deviations from $\Lambda$CDM as strongly. Nonetheless, it is interesting to ask whether similar preferences persist across different probes.

The most visually striking results in Fig.~\ref{fig:individual-datasets} are the $f_{\rm DE}(z)$ reconstructions using BAO data alone. As discussed in Appendix~\ref{sec:mocks}, the $f_{\rm DE}(z)$ reconstructions are affected by a degeneracy between their lowest-redshift amplitude and the Hubble constant. When this degeneracy is not sufficiently constrained by low-redshift measurements, it leads to an unconstrained direction in the likelihood and an apparent upward drift in the $f_{\rm DE}(z)$ constraints. This occurs in the $f_{\rm DE}(z)$ reconstructions from BAO alone, as these datasets have minimal constraining power below $z\lesssim0.3$. This is not a genuine preference of the BAO measurements: we confirmed that the same upward drift appears in reconstructions performed on mock datasets generated from a fiducial $\Lambda$CDM cosmology. Hence, our methodology is currently not well suited for BAO-only $f_{\rm DE}(z)$ reconstructions.

A prominent feature in the $f_{\rm DE}(z)$ reconstructions from most BAO and SNe~Ia combinations is an apparent local maximum between $z\in[0.6,0.8)$ (Sec.~\ref{sec:fde-res}). The reconstructions from Union3.1 and Pantheon+ data show similar preferences at $\sim1$--$2\sigma$ significance, while DES-Dovekie slightly prefers a local maximum in the subsequent redshift bin, $z\in[0.8,1.1)$. At lower redshifts, Union3.1 and Pantheon+ show a marginal preference for $f_{\rm DE}>1$, while DES-Dovekie prefers no strong deviations from the cosmological constant limit.

The $w_{\rm DE}(z)$ results in Fig.~\ref{fig:individual-datasets} help explain the oscillatory features observed in reconstructions from dataset combinations (Sec.~\ref{sec:wde-res}). Initial fluctuations above $w_{\rm DE}=-1$ in the lowest-redshift bin are seen clearly in the SNe~Ia reconstructions, whereas the subsequent rise and fall surrounding $z\in[0.6,0.8)$ are weakly preferred by the BAO datasets, particularly DESI.

\section{\label{sec:full-posteriors}Full Posteriors}

\begin{figure*}
    \centering
    \includegraphics[width=0.9\linewidth]{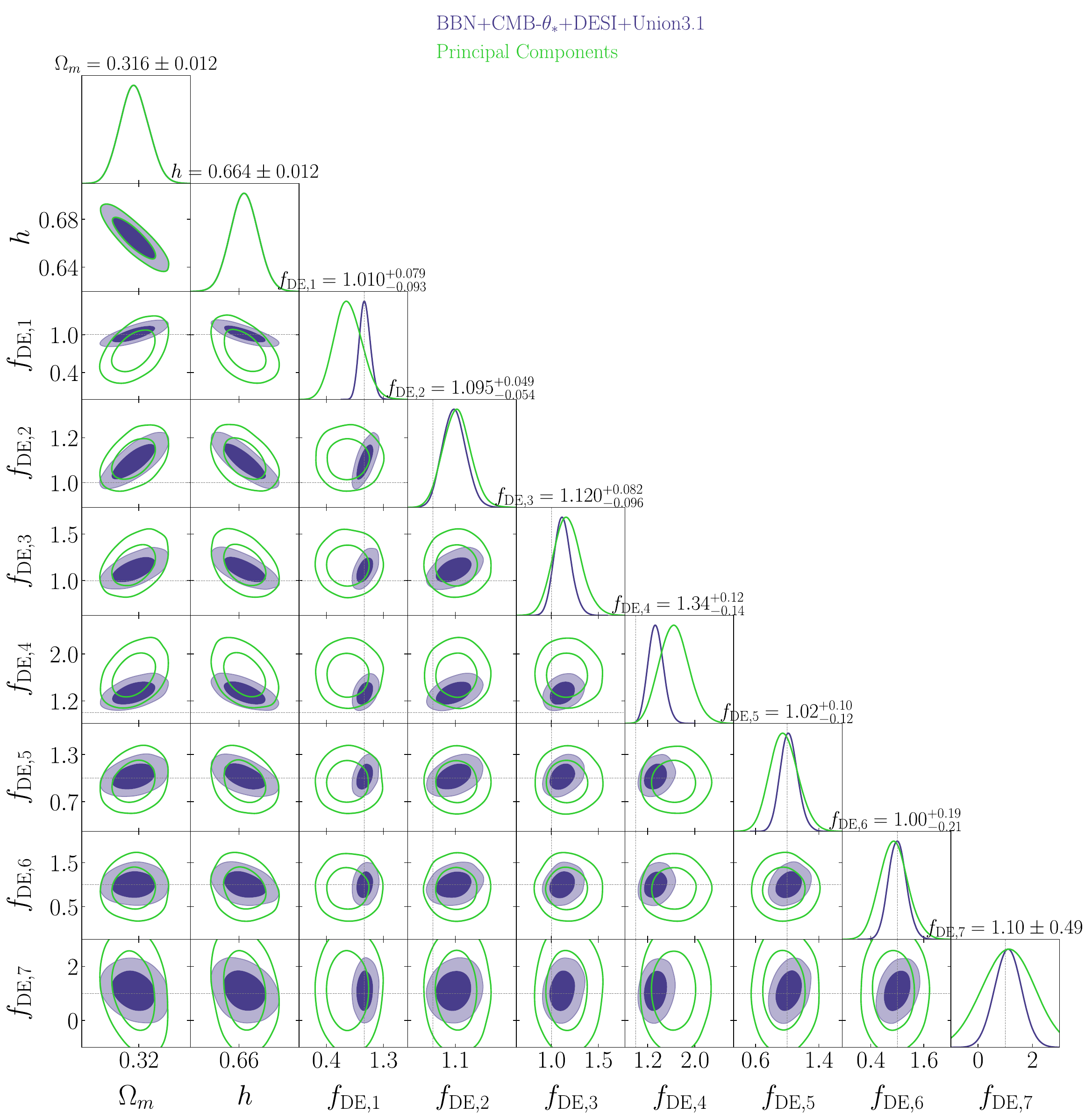}
    \caption{Full marginalized posteriors for the matter density $\Omega_m$, dimensionless Hubble rate $h$, and the $f_{\rm DE}(z)$ bins from the direct reconstruction (blue) and localized principal component analysis (green). The $1\sigma$ parameter limits above the 1D histograms correspond to the direct reconstruction.
    }
    \label{fig:fde-triangle}
\end{figure*}

\begin{figure*}
    \centering
    \includegraphics[width=0.9\linewidth]{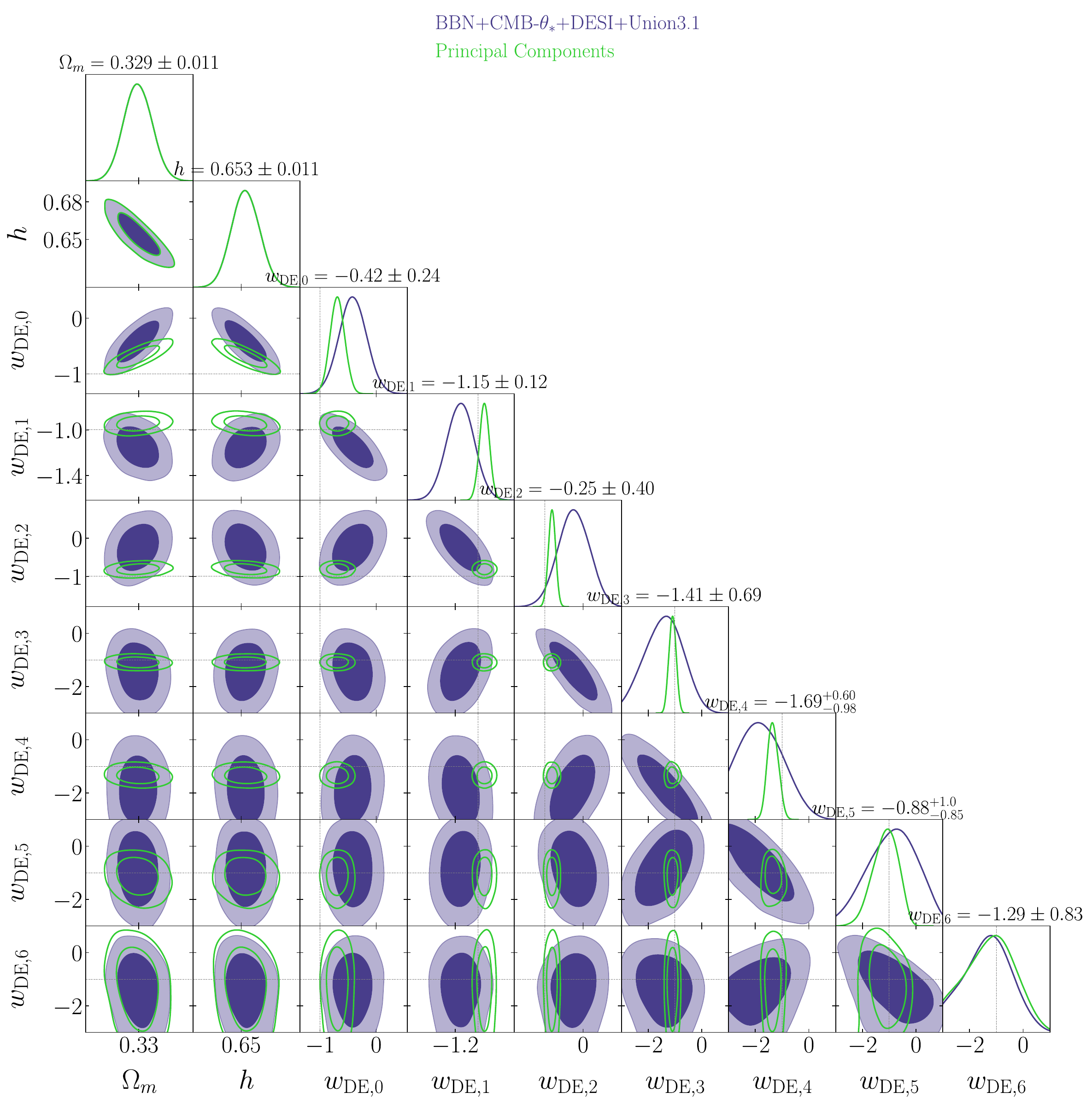}
    \caption{Full marginalized posteriors for the matter density $\Omega_m$, dimensionless Hubble rate $h$, and the $w_{\rm DE}(z)$ bins from the direct reconstruction (blue) and LPCA (green). The $1\sigma$ parameter limits above the 1D histograms correspond to the direct reconstruction. LPCA yields significantly tighter constraints on the amplitude of $w_{\rm DE}(z)$ below $z\lesssim1.1$, at the expense of decreased redshift localization (Fig.~\ref{fig:wde-lpca}).}
    \label{fig:wde-triangle}
\end{figure*}

Figures~\ref{fig:fde-triangle} and \ref{fig:wde-triangle} show the full marginalized posteriors for the cosmological and reconstruction parameters of the $f_{\rm DE}(z)$ and $w_{\rm DE}(z)$ reconstructions, respectively, using a representative dataset combination, DESI+Union3.1. The $f_{\rm DE}(z)$ bins have a uniformly positive degeneracy direction with the Hubble constant, $h$, whereas the $w_{\rm DE}(z)$--$h$ degeneracy direction alternates between bins. This possibly explains why our $f_{\rm DE}(z)$ reconstructions are more susceptible to bias through their degeneracy with $h$, as this degeneracy pulls all $f_{\rm DE}(z)$ amplitudes in the same direction. Also shown in this Figure are the posteriors of the uncorrelated amplitudes obtained from localized principal component analysis (Sec.~\ref{sec:lpca}). For the $w_{\rm DE}(z)$ reconstruction, the LPCA amplitudes have much smaller uncertainties than the directly reconstructed amplitudes, although this comes at the expense of decreased redshift localization; see the redshift weights in Fig.~\ref{fig:wde-lpca}.

\bibliography{biblio}

\end{document}